\documentclass[traditabstract]{aa} 
                               
\usepackage{natbib}                                
\usepackage{graphicx}
\usepackage{txfonts}
\usepackage{longtable}
\newcommand{\HI}{H{\scriptsize I}}
\newcommand{\NGAL}{NGC$~$4330}

\begin{document}

   \title{A Virgo Environmental Survey Tracing Ionised Gas Emission (VESTIGE).}
   \subtitle{II. Constraining the quenching time in the stripped galaxy \NGAL}
   \author{M. Fossati\inst{1,2},  
          J.T. Mendel\inst{1},
          A. Boselli\inst{3,}\thanks{Visiting Astronomer at NRC Herzberg Astronomy and Astrophysics.},  
          J.C. Cuillandre\inst{4},
          B. Vollmer\inst{5},
          S. Boissier\inst{3},
          G. Consolandi\inst{6},
          L. Ferrarese\inst{7},
          S. Gwyn\inst{7}, 
          P. Amram\inst{3},  
          M. Boquien\inst{8}, 
          V. Buat\inst{3},
          D. Burgarella\inst{3},
          L. Cortese\inst{9},   
          P. C{\^o}t{\'e}\inst{7},
          S. C{\^o}t{\'e}\inst{7},
          P. Durrell\inst{10},
          M. Fumagalli\inst{11}, 
          G. Gavazzi\inst{12},
          J. Gomez-Lopez\inst{3},
          G. Hensler\inst{13},
          B. Koribalski\inst{14}, 
          A. Longobardi\inst{15},
          E.W. Peng\inst{15},
          J. Roediger\inst{7},
          M. Sun\inst{16},
          E. Toloba\inst{17}
     }

\institute{     
                Max-Planck-Institut f\"{u}r Extraterrestrische Physik, Giessenbachstrasse, 85748, Garching, Germany 
                \email{mfossati@mpe.mpg.de}
        \and  
                Universit{\"a}ts-Sternwarte M{\"u}nchen, Scheinerstrasse 1, D-81679 M{\"u}nchen, Germany
        \and
                 Aix Marseille Universit\'e, CNRS, LAM, Laboratoire d'Astrophysique de Marseille, Marseille, France
                \email{alessandro.boselli@lam.fr}
        \and
                CEA/IRFU/SAP, Laboratoire AIM Paris-Saclay, CNRS/INSU, Universit\'e Paris Diderot, Observatoire de Paris, PSL Research University, F-91191 Gif-sur-Yvette Cedex, France
        \and
                Observatoire Astronomique de Strasbourg, UMR 7750, 11, rue de l'Universit\'e, 67000, Strasbourg, France
        \and
                INAF Osservatorio Astronomico di Brera, Via Brera 28, 20121 Milano, Italy
        \and
                NRC Herzberg Astronomy and Astrophysics, 5071 West Saanich Road, Victoria, BC, V9E 2E7, Canada
         \and       
                Unidad de Astronom{\'i}a, Fac. de Ciencias Basicas, Universidad de Antofagasta, Avda. U. de Antofagasta 02800, Antofagasta, Chile
         \and    
                International Centre for Radio Astronomy Research, The University of Western Australia, 35 Stirling Highway, Crawley WA 6009, Australia
         \and
                Department of Physics and Astronomy, Youngstown State University, Youngstown, OH 44555, USA
         \and
                Institute for Computational Cosmology and Centre for Extragalactic Astronomy, Department of Physics, Durham University, South Road, Durham DH1 3LE, UK
        \and
               Universit\'a di Milano-Bicocca, Piazza della scienza 3, 20100, Milano, Italy
       \and
               Department of Astrophysics, University of Vienna, Turkenschanzstrasse 17, 1180, Vienna, Austria
        \and
                Australia Telescope National Facility, CSIRO Astronomy and Space Science, P.O. Box 76, Epping, NSW 1710        
        \and        
               Department of Astronomy, Peking University, Beijing 100871, China
        \and
                Department of Physics and Astronomy, University of Alabama in Huntsville, Huntsville, AL 35899, USA
        \and
               Department of Physics, University of the Pacific, 3601 Pacific Avenue, Stockton, CA 95211, USA
              }

\authorrunning{M. Fossati et al.}
\titlerunning{VESTIGE II. Quenching times in \NGAL}

\date{}
 
\abstract  
{The Virgo Environmental Survey Tracing Ionised Gas Emission (VESTIGE) is a blind narrow-band H$\alpha$+[NII]
imaging survey carried out with MegaCam at the Canada-France-Hawaii Telescope. During pilot observations taken
in the spring of 2016 we  observed \NGAL, an intermediate mass ($M_* \simeq 10^{9.8} \rm{M_\odot}$) edge-on star 
forming spiral currently falling into the core of the Virgo cluster. While previous H$\alpha$ observations showed a 
clumpy complex of ionised gas knots outside the galaxy  disc, new deep observations revealed a low surface brightness 
$\sim 10$ kpc tail exhibiting a peculiar filamentary structure. The filaments are remarkably parallel to one another and clearly 
indicate the direction of motion of the galaxy in the Virgo potential. Motivated by the detection of these features which indicate  
ongoing gas stripping, we collected literature photometry in 15 bands from the far-UV to the far-IR and deep optical 
long-slit spectroscopy using the FORS2 instrument at the ESO Very Large Telescope. Using a newly developed Monte 
Carlo code that jointly fits spectroscopy and photometry,
we reconstructed the star formation histories in apertures along the major axis of the galaxy. Our results have been validated
against the output of CIGALE, a fitting code which has been previously used for similar studies.
We found a clear outside-in gradient with radius of the time when the quenching event started: the outermost radii   
were stripped $\sim 500$ Myr ago, while the stripping  reached the inner 5 kpc from the centre  in the last 100 Myr.
Regions at even smaller radii are currently still forming stars fueled by the presence of \HI\ and H$_2$ gas. 
When compared to statistical studies of the quenching timescales in the
local Universe we find that ram pressure stripping of the cold gas is an effective mechanism to reduce the transformation 
times for galaxies falling into massive clusters. Future systematic studies of all the active galaxies observed by VESTIGE in the Virgo cluster
will extend these results to a robust statistical framework.  }


   \maketitle
%

\section{Introduction}
It has long been known that galaxies are not uniformly distributed in the Universe.  Works 
by e.g. \citet{Oemler74}, \citet{Dressler80}, and \citet{Balogh97} showed that galaxies in high-density environments are 
preferentially red and form new stars at a lower rate compared to similar objects in less dense environments. 
The more recent advent of large-scale photometric and spectroscopic surveys have confirmed with large statistics 
that the most massive dark matter haloes (cluster of galaxies) are mainly composed of quiescent objects 
with an elliptical or lenticular shape. Conversely star forming  disc-like systems are dominant in the field 
and in lower mass haloes \citep{Balogh04, Kauffmann04, Baldry06}. Moreover, gas-rich galaxies are found in 
lower density environments than optically selected galaxies are \citep{Koribalski04, Meyer04}, suggesting 
a strong interplay between the gas cycle and the star formation activity of galaxies. 

Different mechanisms have been proposed in the literature to explain this observational evidence (e.g. 
gravitational and hydrodynamical interactions), as reviewed in \citet{Boselli06, Boselli14c}. The role and the 
importance of each of these processes in shaping galaxy evolution, however, is still under debate. 
Many variables are indeed at play, including the mass of the perturbed galaxy, the mass and density of 
the perturbing region, and the epoch of the interaction.

Theoretical works and numerical simulations have shown that once galaxies fall into a more massive halo and start to orbit in it as satellites, 
they lose the ability to accrete fresh gas from the cosmic web \citep[see e.g.][]{Larson80, Dekel06}. This leads 
to a quenching of the star formation activity once the cold gas located on the disc is fully transformed into stars, 
which occurs on timescales of several Gyr. This phenomenon, variously called 
`starvation', `strangulation'  \citep{Larson80}, or `overconsumption'  \citep{McGee14}, 
is supported by a statistical analysis of a large sample of galaxies from the Sloan Digital Sky Survey (SDSS),
which suggests a quenching timescale of $\sim$ 5--7 Gyr since their first accretion as satellites in group-like 
environments \citep{Wetzel13, Hirschmann14, Fossati17}. 
These authors further broke the quenching event into two phases: a delay time ($\sim 2-5$ Gyr)
during which the star formation activity of the satellite is unaffected.  This phase is characterised by
the replenishment of the molecular gas reservoir via gas cooling from a warmer phase (e.g. atomic).
After this time, the star formation rate (SFR) rapidly fades due to the reduction 
of the molecular gas mass via star formation.

However, for satellite galaxies in massive clusters of galaxies, several 
authors have found shorter ($1-2$ Gyr) quenching timescales \citep{Boselli16a, Oman16}.
Although the different methods make the comparison to the delayed+rapid quenching model
non-trivial, it is generally found that both the delay and the fading phases get shorter once
the effect of quenching in haloes different from the cluster environment are taken into account
\citep{Oman16}. However, other authors have found that at least a fraction of the cluster satellite population 
is gradually and continuously quenched on timescales of $\sim 2-5$ Gyr \citep{Haines15, Paccagnella16}.

These results point to ram pressure as a compelling process to explain the high 
passive fractions and shorter quenching timescales in cluster environments. 
Recent hydrodynamical simulations of individual galaxies show that ram pressure is an efficient gas 
stripping and quenching mechanism up to $\sim$ 1 virial radius of the cluster \citep{Tonnesen09}. 
Furthermore, models indicate that the cold gas component of the interstellar medium (ISM) is stripped outside-in, 
forming truncated discs with bent shapes in the z-plane and long tails of stripped gas. This phenomenon happens
already in the very early phases of the stripping process, on timescales of 20 -- 200 Myr \citep{Roediger05}. 
The recent observation of several late-type galaxies with long tails of gas without associated streams of old stars
at large cluster-centric distances corroborate this scenario \citep[e.g.][]{Yagi10, Scott12, Fossati12, Fossati16}. 
These observations have shown that most of the gas, in particular that loosely bound to the 
potential of the galaxy in the outer disc, is stripped on very short timescales \citep[$\sim$ 100--200 Myr;][]{Boselli06a}. 
Detailed studies of individual objects in clusters are therefore fully complementary to statistical studies for 
understanding the physics of the mechanisms operating in clusters of galaxies.
However, only a few studies attempted a reconstruction of the age of the stellar populations in local 
ram pressure stripped galaxies \citep[see e.g.][]{Crowl08, Pappalardo10, Merluzzi16}. More recently, \citet{Fritz17}
has presented an analysis of the star formation histories in a ram pressure stripped galaxy observed with the 
MUSE integral field unit from the GAs Stripping Phenomena (GASP) survey. This galaxy has a truncated ionised gas 
disc typical of stripped objects. However, their non-parametric fitting code has a limited age resolution (two age bins 
in the last 500 Myr), and is therefore  not sensitive to recent and rapid quenching events.
In summary, the variable quality of the datasets and of the analysis method has so far hampered a complete, spatially 
resolved view of how fast the quenching proceeds at different  galactocentric radii. 

\begin{figure*}
\centering
\includegraphics[width=16.5cm]{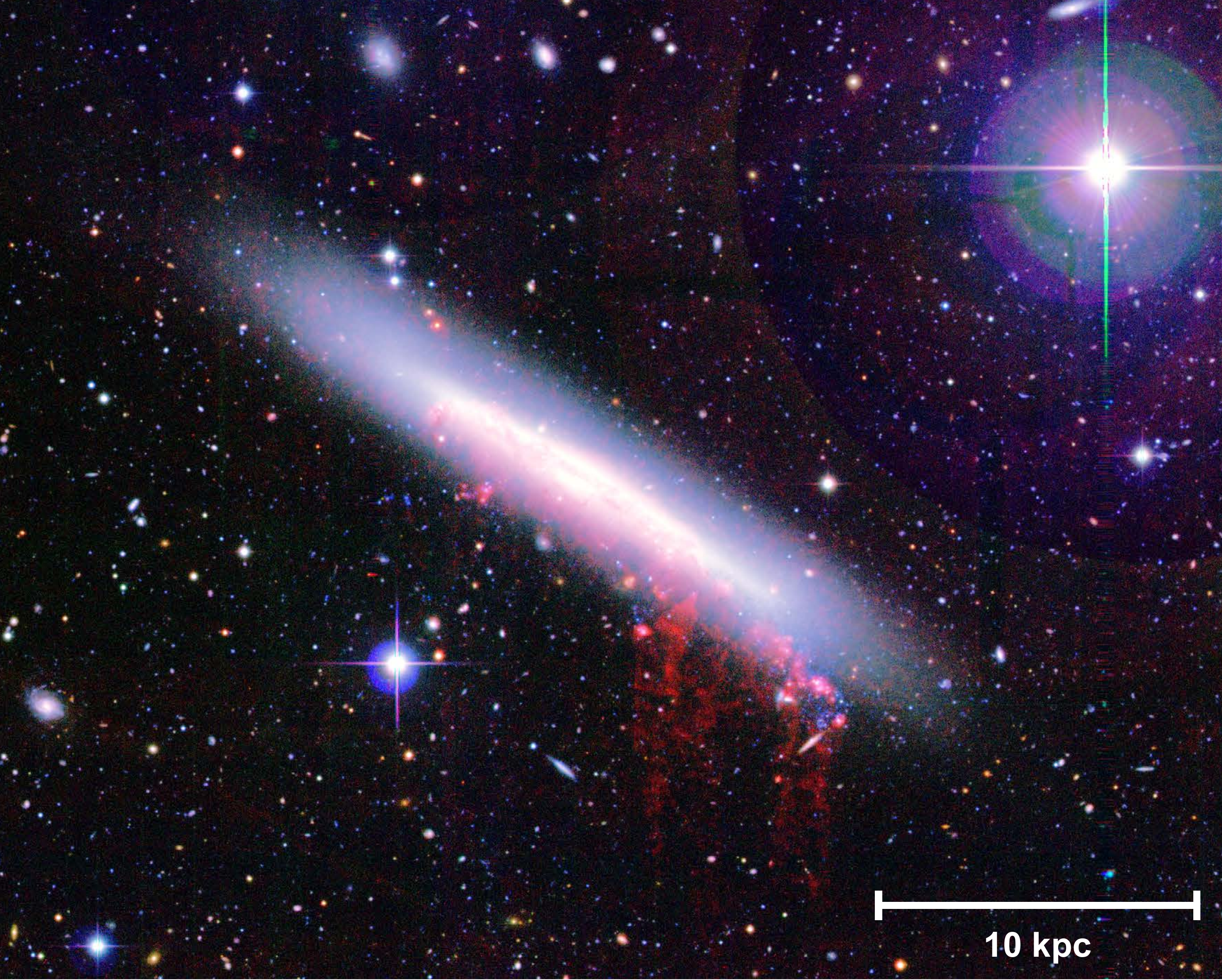}
\caption{Pseudo-colour image of \NGAL\ obtained from CFHT NGVS broad-band and VESTIGE narrow-band images. 
The H$\alpha$ net emission is shown as the red channel of the image. A faint tail of ionised gas trails downstream 
of the galaxy while it is travelling towards the NW. The image is oriented such that north is up and 
east is left. A physical scale of 10 kpc is shown by the white segment.}
\label{N4330_colorima}
\end{figure*}

In this work we present the analysis of the quenching timescales in \NGAL, a spiral edge-on
galaxy in the Virgo cluster. Deep H$\alpha$ narrow-band (NB) 
observations have been recently taken for this galaxy as part 
of pilot observations for the Virgo Environmental Survey Tracing Ionized Gas Emission (VESTIGE; P.I. A. Boselli).
VESTIGE is a large programme which has been allocated 50 nights of observing time at the Canada-France-Hawaii 
Telescope (CFHT) with the aim of covering the central 104 deg$^2$ of the Virgo cluster (up to 1 $r_{\rm vir}$) with sensitive
H$\alpha$ and $r$-band observations. We refer the reader to \citet[][Paper I]{Boselli18} for a full description of the survey.
These observations have revealed a low surface brightness, extended, one-sided tail of ionised gas being removed from the galaxy disc.
This feature, combined with the truncated H$\alpha$, \HI, and CO morphologies compared to the extent of the stellar disc, 
unambiguously points to ram pressure stripping as the mechanism responsible for the quenching of the star formation 
activity in \NGAL\ \citep{Chung07, Abramson11, Lee17}. 
That the quenching process is still ongoing, coupled with the edge-on morphology, made this galaxy an ideal candidate 
for studies of the quenching timescale as a function of the galactocentric radius. The first attempt to reconstruct the radial truncation 
of the star formation history of \NGAL\ was presented by \citet{Abramson11}. These authors used a combination of UV and optical colours, coupled with stellar 
population models to derive an estimate of the quenching times out to 8 kpc from the galaxy centre. 
In this work, we extend their analysis by exploiting the excellent multiwavelength photometric coverage of the Virgo cluster 
and using high spatial resolution data from the far-UV (FUV) to the far-IR (FIR). Furthermore, we complemented
this dataset with deep medium-resolution optical spectroscopy to obtain the best constraints on the quenching times
in the periphery of the disc (out to 10 kpc) where the star formation activity has ceased completely. 

In this paper we use a novel spectrophotometric fitting method 
which uses Monte Carlo techniques to sample the parameter space. This method is coupled with state-of-the-art 
models for the growth and quenching of disc galaxies as a function of radius. We test our new code by comparing it with
the results from CIGALE \citep{Noll09}, a fitting code which has  already been 
used to characterise the quenching histories of galaxies in cluster environments \citep{Boselli16a}. 
In section \ref{sec_galaxy}, we present the target galaxy and the photometric and spectroscopic data we use in this paper, 
and in section \ref{sec_tail} we describe the properties of the H$\alpha$ tail discovered by the VESTIGE observations. We 
introduce the codes and the models used to reconstruct the star formation histories in section \ref{sec_fitting}, and we present
our results in section \ref{sec_results}. In section \ref{sec_discussion} we discuss those results in the context of a ram pressure stripping scenario
in local galaxy clusters and we summarise our conclusions in section \ref{sec_conclusions}.

Similarly to other papers in this series, magnitudes are given in the AB system \citep{Oke74} and we 
assume a flat $\Lambda$CDM Universe with $\Omega_M = 0.3$, $\Omega_\Lambda = 0.7$, and 
$H_0 = 70~\rm{km~s^{-1}~Mpc^{-1}}$. With the adopted cosmology and assuming a 
distance of 16.5 Mpc to the Virgo cluster \citep{Gavazzi99, Mei07, Blakeslee09}, 1$"$ on the sky corresponds to a physical scale of 80 pc. Where necessary,
we adopt a \citet{Chabrier03} initial mass function.

\begin{figure*}
\centering
\includegraphics[width=16.5cm]{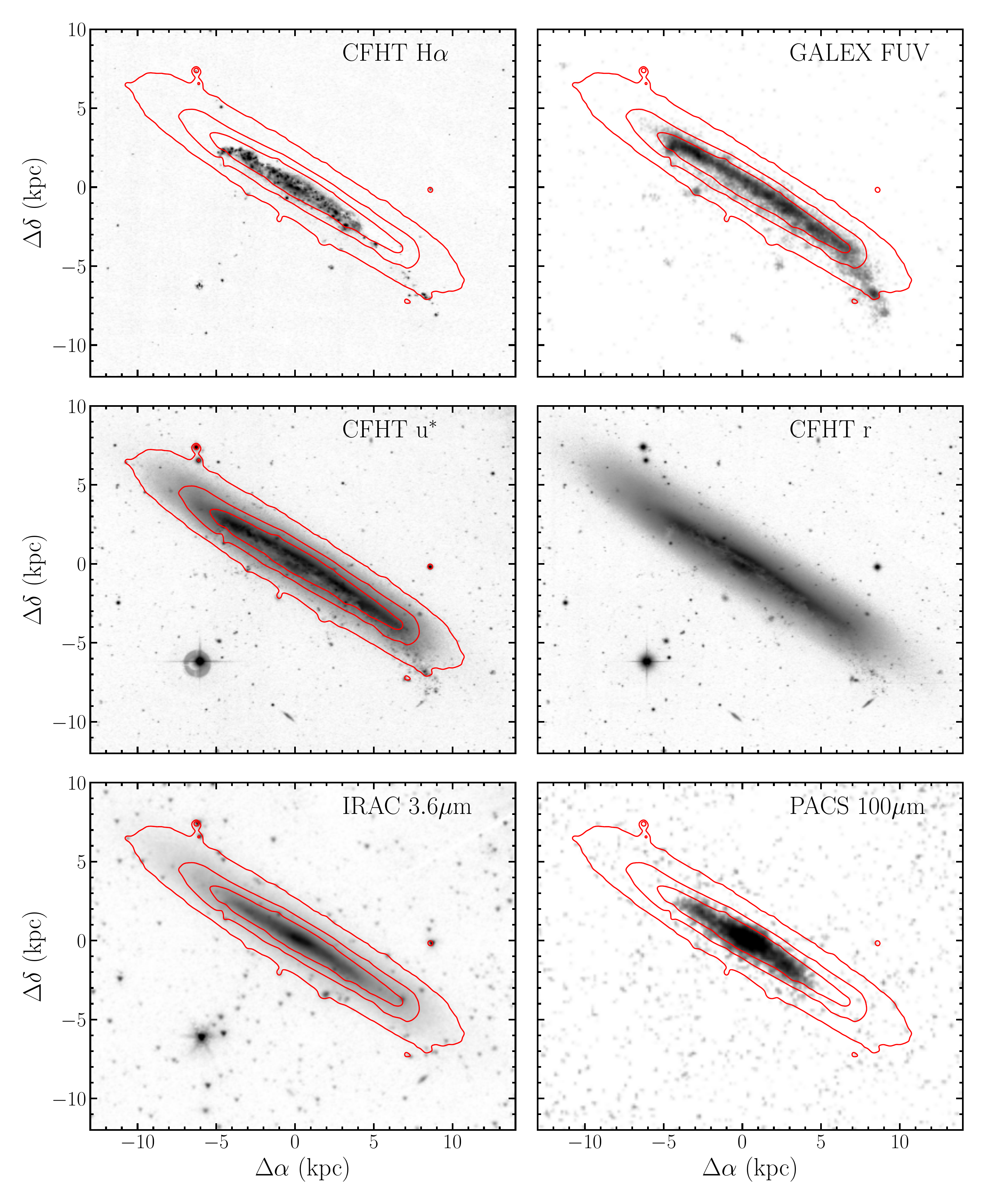}
\caption{Multifrequency images of \NGAL\ at their intrinsic resolution. From the top left to the bottom right: CFHT/VESTIGE H$\alpha$, 
GALEX FUV, CFHT/NGVS $u^*$, CFHT/VESTIGE $r$, {\it Spitzer}/IRAC 3.6 $\mu$m, and {\it Herschel}/PACS 100 $\mu$m. All the images are on the
same scale and are oriented such that north is up and east is left. The red contours are drawn from the $r$-band image at the $22{\rm{nd}}$, $23{\rm{rd}}$, and 25${\rm{th}}$ 
mag$/\rm{arcsec}^2$ and clearly highlight the different amount of disc truncation at different wavelengths. The presence of a tail extending to the SW is clear
in the FUV image.}
\label{N4330_multiwave}
\end{figure*}

\section{\NGAL}  \label{sec_galaxy}
\NGAL\ (VCC 630) is a highly inclined (84 deg), intermediate mass \citep[$M_* \simeq 10^{9.7} \rm{M_\odot}$,][]{Boselli16a} Scd galaxy in the Virgo cluster 
located at a projected distance of $\simeq$ 600 kpc (0.4 $\times~r_{\rm vir}$) from M87. 
Spatially resolved radio observations made with the Very Large Array in C+D configuration have revealed 
the presence of a long tail of \HI\ gas \citep{Chung07} on the trailing side
of the interaction, pointing in an opposite direction with respect to M87. The tail also hosts some young star forming regions
(visible as the blue knots in Figure \ref{N4330_colorima}) which were first noted in the optical, UV, and H$\alpha+\rm{[NII]}$\footnote{
Hereafter, we  refer to the net H$\alpha+\rm{[NII]}$ image simply as H$\alpha$, 
unless otherwise stated.} images by \citet{Abramson11}, and later confirmed in deeper GALEX/FUV and optical NGVS images by \citet{Boissier12}.
Similar features have also been found  in other ram pressure stripped tails \citep{Fumagalli11, Fossati12, Fumagalli14}, although 
the physical conditions leading to their presence or absence in stripped tails are still poorly understood.
Another peculiar morphological feature is represented by a truncated and asymmetric disc in the young 
stellar populations revealed by H$\alpha$ and UV images, compared to the more extended and symmetric 
disc seen in older stellar populations \citep{Abramson11}. 
Figure \ref{N4330_multiwave} presents images of \NGAL\ in six photometric bands from the FUV to the FIR. The details
of the observations are described in Section \ref{Vestige_phot} and \ref{Ancillary_phot}. The red contours are drawn from
the $r$-band image at the $22{\rm{nd}}$, $23{\rm{rd}}$, and 25${\rm{th}}$ mag$/\rm{arcsec}^2$ and clearly highlight the different
extension of the disc at different wavelengths. It is interesting to note how the dust disc (as traced by the {\it Herschel}/PACS
map) is truncated at the same radius as the youngest stars (traced by the H$\alpha$ image), as first found by \citet{cortese10}.

Recently \citet{Lee17} presented high-resolution CO maps of \NGAL\ revealing an asymmetric CO morphology
which closely coincides with the location of the H$\alpha$ emission. However, the UV emission is more extended than 
the molecular gas distribution, which suggests a recent and rapid quenching of the star formation activity occurred during the
stripping of all the gaseous components from the outer regions of NGC4330.
 
\citet{Vollmer12} presented a ram pressure stripping model of NGC4330 tailored to reproduce the observed 
morphology and extension of different stellar populations. This model suggests that the stripping occurs almost face-on 
(the angle between the galaxy's disc and the wind direction is $75\deg$) and the peak of the stripping force has not yet been reached.
The best fit to the observations is  obtained $\sim 100$ Myr before the peak of stripping is reached. This explains the 
significant star formation activity in the galaxy centre where the quenching has not yet been effective.

\subsection{VESTIGE photometry} \label{Vestige_phot}
In order to detect low surface brightness features in the H$\alpha$ emission, we selected NGC4330 as one of the highest priority 
targets for VESTIGE pilot observations. The observations were carried out in April 2016 using MegaCam at the CFHT with the NB filter (MP9603) and the  broad-band $r$ filter (MP9602). The observing strategy is described in detail in \citet{Boselli16} and in Paper I. 
In brief, we observed a sequence of seven offset pointings where the large dithers are ideal for minimising the reflection of bright stars
in the stacked images. The exposure times of individual  NB and $r$-band frames were 660s and 66s respectively. Each sequence
was repeated three times for total integration times of 13860s and 4620s in the two filters, respectively. The observations were taken 
in excellent weather conditions under photometric skies and with a median seeing FWHM $\simeq$ 0.62"  and 0.65" in the NB and $r$-band filter, respectively.

The data have been reduced using the Elixir-LSB package \citep{Ferrarese12}, a data reduction pipeline optimised to enhance the
signal-to-noise ratio of extended low surface brightness features by removing the instrumental background and scattered light from the science frames.
This is obtained by jointly reducing a sequence of concatenated science exposures, which we obtained using 
specifically designed dither patterns. 
The photometric zero points were tied to 
Pan-STARRS photometry for both filters with a final photometric uncertainty of $\sim 2-3 \%$ (see Paper I for the colour transformations used). 
The image stacking and astrometric 
registration are performed using the MegaPipe procedures \citep{Gwyn08}.

The H$\alpha$ image is obtained by 
subtracting the stellar continuum from the NB image, thus leaving only the nebular line contribution. The details and calibrations of this procedure
will be described in a forthcoming paper (Fossati et al. in preparation). Here we briefly summarise the main steps. The continuum image is obtained by scaling the
$r$-band image to take into account the difference in central wavelength of the two filters. Because the slope of the stellar continuum spectrum
can be estimated from an optical colour we include a colour-term ($g-r$) in the transformation, which we calibrated extracting synthetic magnitudes from 
the spectra of 50,000 stars taken from SDSS. For each pixel we apply the  equation
\begin{equation}
NB_{\rm cont} = m_r - 0.1713 \times (m_g - m_r) + 0.0717
,\end{equation}
where all the variables are expressed in AB magnitudes in the CFHT filters. To limit the surface brightness uncertainty due to the 
subtraction of the images we apply a median 3x3 pixel smoothing in the $NB_{\rm cont}$ image for all the pixels which have a signal-to-noise 
ratio below 10. The native spatial resolution is kept in the other pixels.  Lastly, we subtract the $NB_{\rm cont}$ image from the NB image and we
multiply the pixel monochromatic flux densities by the filter width (106 \AA) in order to obtain the values of the line flux per pixel. 
Although in principle the $r$-band flux includes the line emission, this is dominant only in regions with high line equivalent width (EW$>$500). 
Using a sample of SDSS spectra we derived an iterative method to correct for this effect. However, for this galaxy the H$\alpha$ EW
is relatively high only in the tail region where the emission line flux does not show any underlying continuum. There, the H$\alpha$ EW
is highly uncertain due to the faintness of the emission and the non-detection in the $r$-band filter. For this reason, and after testing that
the derived H$\alpha$ flux does not change if the correction is applied, we decided not to apply any correction.
 
\subsection{Ancillary photometry} \label{Ancillary_phot} 
NGC4330 has been observed across the electromagnetic spectrum in 20 bands from the FUV to the FIR. 
Deep FUV and near-UV (NUV) observations were taken with the Galaxy Evolution Explorer \citep[GALEX, ][]{Martin05} 
in 2007 (Programme 79, Cycle 1) with an exposure time of 17000s in both bands. The data were processed with the
GALEX pipeline and were downloaded from the database of the GALEX Ultraviolet Virgo Cluster Survey \citep[GUViCS,][]{Boselli11a}. 
The typical spatial resolution element is FWHM $\sim 5"$.

Optical images in the $u, g, i, z$ broad bands were taken at CFHT as part of the Next Generation Virgo 
Cluster Survey \citep[NGVS, ][]{Ferrarese12}. The data were reduced with the Elixir-LSB pipeline and the photometric
zero points were tied to SDSS photometry, as was done for the VESTIGE data. The typical FWHM is $\sim 0.55"$ in the $i$ band
and $\sim 0.8"$ in the other bands.

Near-infrared H-band observations were taken at the Calar Alto 2.2m 
telescope by \citet{Boselli00} with a typical FWHM of $\sim 2"$. We re-calibrated the images by tying the flux counts to 
2MASS photometry converted to AB magnitudes. 

Deep {\it Spitzer} \citep{Werner04} images were taken with the IRAC \citep{Fazio04} instrument at 3.6, 4.5, 5.8, 8.0 $\mu$m and with the 
MIPS \citep{Rieke04} instrument at 24 and 70 $\mu$m during the cryogenic mission (PI J.Kenney, Programme 30945). 
The data were downloaded from the {\it Spitzer} Heritage Archive. Because multiple observations were available, we resampled 
and combined them onto a common grid using the Swarp software \citep{Bertin02}. The mean FWHM of IRAC observations ranges from
$1.6"$ to $2.0"$ from channel 1 to 4, while MIPS 24 $\mu$m images have an image quality of $\sim 6"$. We do not use 
MIPS 70 $\mu$m images in our analysis because of their poor spatial resolution (${\rm FWHM} \sim 18"$).

Far-infrared observations of NGC4330 were carried out by the {\it Herschel} Virgo Cluster Survey \citep[HeViCS][]{Davies10}, 
a programme that covered $\sim 60~\rm{deg}^2$ of the Virgo cluster using the PACS \citep{poglitsch10} instrument 
at 100 and 160 $\mu$m, and the SPIRE \citep{griffin10} instrument at 250, 350, and 500 $\mu$m. These data were 
integrated into the {\it Herschel} Reference Survey \citep[HRS,][]{Boselli10}, and the data reduction was carried out as described 
in \citet{Ciesla12} and \citet{Cortese14}. 
The FWHM of PACS observations is $7"$ and $\sim 12"$ at 100 and 160 $\mu$m, respectively, while the FWHM of SPIRE
observations is $\sim 18"$, $\sim 25"$, and $\sim 36"$ at 250, 350, and 500 $\mu$m, respectively. Because of the poorer spatial 
resolution of PACS 160 $\mu$m and of the three SPIRE bands, we do not use these data in this work. In conclusion, 15 photometric bands 
contribute to our dataset, each selected for excellent depth and image quality.

\begin{figure*}
\centering
\includegraphics[width=17.5cm]{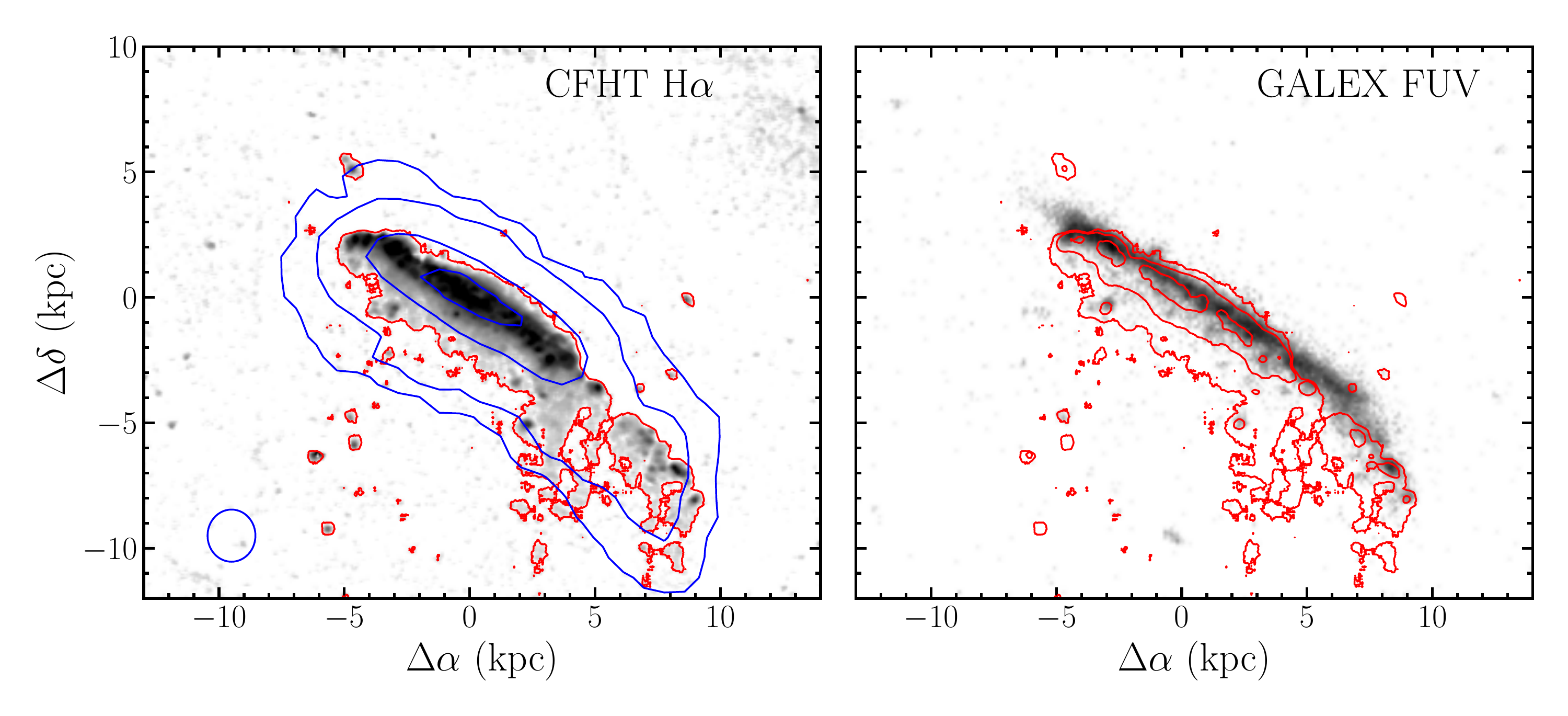}
\caption{Left panel: CFHT/VESTIGE H$\alpha$ image of NGC4330 smoothed at $\sim 2.5"$ resolution. The colour cuts are selected to 
highlight the low surface brightness features. The red contour is at $3\times 10^{-18}~{\rm erg~cm^{-2}~s^{-1}~arcsec^{-2}}$. 
The blue contours are from \HI\ observations obtained with the Very Large Array in C+D configuration, with a beam size (shown in the lower left corner) of $\sim 2$ kpc \citep{Chung09}. 
The contour levels are at $\Sigma$(\HI) = 2, 16, 128, 256 $\times 10^{19}~\rm{cm^{-2}}$.
Right panel:  GALEX FUV image of NGC4330 with H$\alpha$ contours at $3, ~15, ~125~\times 10^{-18}~{\rm erg~cm^{-2}~s^{-1}~arcsec^{-2}}$. 
The H$\alpha$ disc is truncated to a smaller radius compared to the FUV emission, while the H$\alpha$ extraplanar emission is more extended to the S. }
\label{N4330_tail}
\end{figure*}

\subsection{Long-slit spectroscopy} 
In order to derive reliable quenching times in the low surface brightness outskirts of NGC4330, it is essential to couple photometry with
optical spectroscopy covering from the Balmer break to the H$\alpha$ line. The edge-on morphology of NGC4330 makes it ideal for 
long-slit spectroscopy observations. We obtained director discretionary time observations under programme 298.B-5018A (PI A. Boselli) 
with the FOcal Reducer and low dispersion Spectrograph \citep[FORS2; ][]{Appenzeller98} mounted on UT1 of the ESO Very Large Telescope. 

\begin{figure*}
\centering
\includegraphics[width=17.5cm]{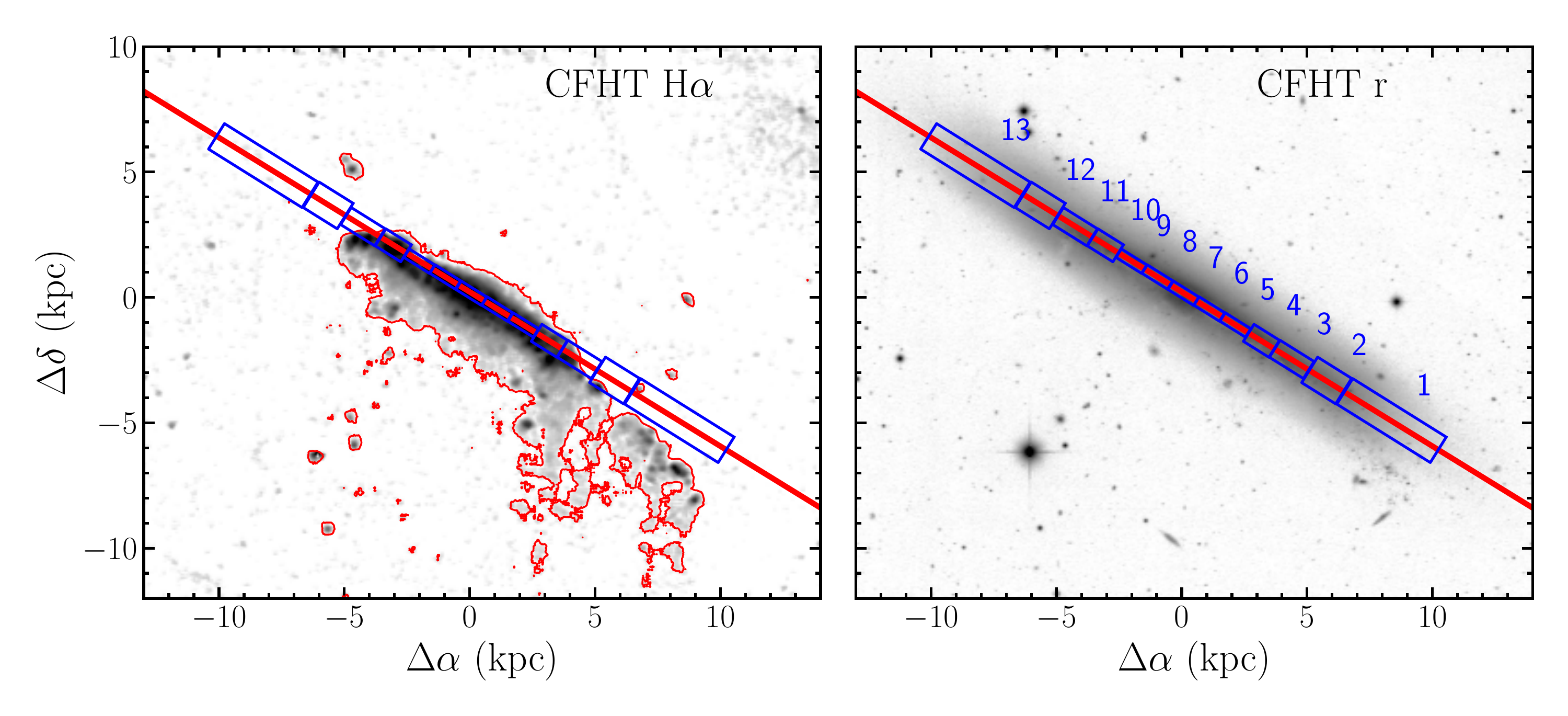}
\caption{Left panel: CFHT/VESTIGE H$\alpha$ image of NGC4330 smoothed at $\sim 2.5"$ resolution. 
The red contour is at $3\times 10^{-18} {\rm erg~cm^{-2}~s^{-1}~arcsec^{-2}}$. Right panel: CFHT/VESTIGE $r$-band image of NGC4330. 
In both panels we highlight the position and the extension of the VLT/FORS2 long slit in red, and the regions in which we performed the aperture 
photometry in blue. In the right panel we also show their ID, as defined in the text. The photometric regions 
have a width along the minor axis of 5$"$ in the inner part of the galaxy in order to minimise the 
mismatch with the spectrum due to differential stellar populations or dust extinction. Going towards the outskirts of the galactic disc we increase
the width to 10$"$, and 15$"$ to preserve the signal-to-noise ratio with decreasing stellar surface brightness.}
\label{N4330_regions}
\end{figure*}

NGC4330 was observed in long-slit spectroscopy (LSS) mode on 26-27 January 2017. 
Observations were conducted under clear skies and in good seeing conditions (FWHM = $0.7"$).  
The slit was aligned along the photometric major axis (Figure \ref{N4330_regions},  red rectangle). With a length of 6.8$'$, the FORS2 slit is 
ideal for covering the full extent of the galaxy. We used a slit aperture of 1.3$"$ to optimise the covered area without 
compromising the spectral resolution. We obtained six exposures of 900s each using the {\sc 600B+22} grism 
which covers the wavelength range 3300 -- 6210 $\AA$, and three exposures of 900s each using the {\sc 600RI+19} grism 
covering the wavelength range 5120 -- 8450 $\AA$. Both grisms have a resolution at the central wavelength $R \sim 1000$. 
The longer exposure time in the blue grism arises from our need to obtain a good signal-to-noise ratio at $\lambda < 4500\AA$ despite the use 
of the red-optimised MIT detector mosaic, which is the only one available for Service Mode observations. 

The raw frames were reduced with the ESO/FORS2 pipeline (v5.5.1), which includes bias subtraction, spectroscopic flat fielding, 
and flux and wavelength calibration. Flux calibration curves were derived from twilight observations of the spectrophotomeric standards Feige66 
and GD108 in both grisms. To reduce the flux calibration uncertainty which can arise from a variety of sources (e.g. sub-optimal 
centring of the star in the $5"$ slit aperture, differential sky extinction between standard and science observations), we averaged
the response curves obtained from each standard star. Because the FORS2 pipeline is not optimised for very extended objects we 
did not use it to subtract the sky background. Instead we carefully selected two regions on both sides of the slit, at 16 kpc from the galaxy centre 
where the galaxy flux is negligible (from the inspection of the deep NGVS $g$ and $i$ images). We used the IRAF Background task to subtract 
a sky spectrum from each spatial element in the 2D spectrum. Lastly, we resampled each 2D spectrum to a common
grid (in pixels of 1.25$\AA$ and 0.25 arcsec) and we combined the single exposures from both grisms using median statistics to 
optimally suppress cosmic rays. We describe in Section \ref{aperturephot} how 1D spectra are extracted along the slit.

\begin{figure}
\centering
\includegraphics[width=8.5cm]{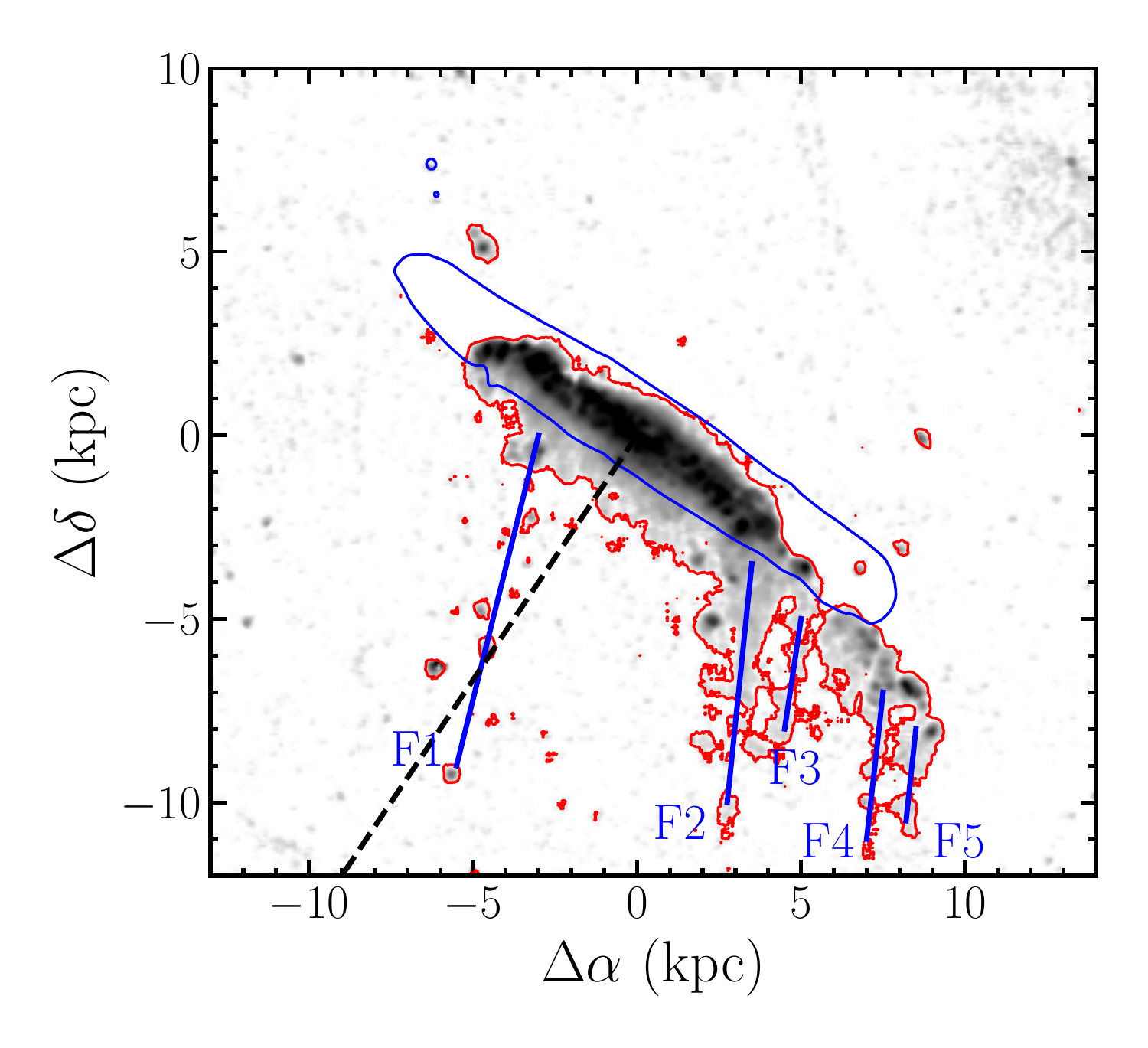}
\caption{CFHT/VESTIGE H$\alpha$ image of NGC4330  shown in Figure \ref{N4330_tail}.  
The blue contour is drawn from the $r$-band image at the $23{\rm{rd}}$ mag$/\rm{arcsec}^2$ and identifies
the region of the galaxy disc. In the text we refer to the H$\alpha$ flux inside this region as the disc flux, 
and we refer to the H$\alpha$ flux outside this region as the stripped flux in the tail. The black dashed line
marks the separation between the upturn region (E of the centre) and the \HI\ tail region (W of the centre).
Five filaments of stripped gas are highlighted with blue segments and labelled F1 to F5 from east to west.}
\label{N4330_filaments}
\end{figure}

\section{The H$\alpha$ tail} \label{sec_tail}
Previous shallow H$\alpha$ observations \citep{Abramson11} revealed a bent and strongly truncated H$\alpha$ disc with 
high surface brightness knots extending in the south-west direction and suggestive of an ionised gas tail. 
The depth of our VESTIGE imaging observations revealed such a tail extending up to 10 kpc (projected distance) from the galaxy disc.
This feature is faint, and has a typical surface brightness of $(3-5) \times 10^{-18} {\rm erg~cm^{-2}~s^{-1}~arcsec^{-2}}$. 
We highlight the low surface brightness H$\alpha$ features in Figure \ref{N4330_tail} (left panel). 

The \HI\ contours shown in Figure \ref{N4330_tail} suggest that the H$\alpha$ has a similar extension to that of  the \HI\ tail (which possibly extends further to the S and W). However, at the $\sim 2$ kpc resolution of \HI\ observations we conservatively
note that the extension of the tails in these two gas phases is strikingly similar. The right panel of Figure \ref{N4330_tail} shows the FUV emission
from young stars with superimposed H$\alpha$ contours. As already noted in Figure \ref{N4330_multiwave}, the H$\alpha$ disc is more
truncated along the galaxy major axis than the FUV disc, a feature  generally interpreted as an indication of recent
quenching of the star formation activity \citep{Boselli06a, Abramson11}. Moreover, we note that even where we find extraplanar FUV emission (in the \HI\ tail region) this is more
concentrated in the N direction than the ionised gas which extends downstream towards the south. The tail hosts recent massive star formation
from which ionised gas is stripped further south. 

To quantify the amount of stripped ionised gas we divide the galaxy in two regions. The one associated with the stellar disc is defined as being inside the
blue contour shown in Figure \ref{N4330_filaments}, and it is drawn from the $r$-band image at the $23{\rm{rd}}$ mag$/\rm{arcsec}^2$. This choice, 
although arbitrary, corresponds to the high surface brightness H$\alpha$ emission and covers the galaxy disc without extending to the faint stellar halo
visible in Figure \ref{N4330_colorima}. The H$\alpha$ flux outside this region is though to  come from the stripped gas. Inside the disc we 
measure $f(\rm{H\alpha+[NII]})_{\rm disc} = (2.7 \pm 0.1) \times 10^{-13} {\rm erg~s^{-1}}$, while in the tail we measure 
$f(\rm{H\alpha+[NII]})_{\rm tail} = (0.53 \pm 0.21) \times 10^{-13} {\rm erg~s^{-1}}$. The flux in the tail accounts for 16\% of the total flux, 
and is dominated by a few bright star forming knots.  

The H$\alpha$ tail is clearly more prominent in the direction of the \HI\ tail; we quantify this asymmetry by separately computing the  $\rm{H\alpha+[NII]}$
flux to the E and the W of the black dashed line in Figure \ref{N4330_filaments}. We find that the flux on the eastern side of the tail is 1/5 of the flux on 
the other side. Using numerical simulations, \citet{Vollmer12} interpreted this asymmetry as  evidence that the galaxy is
not travelling face-on through the wind; a small deviation ($\sim 15$ deg) from a face-on wind can cause the observed H$\alpha$ spatial distribution.

It is interesting to note that the tail morphology, while  compact immediately downstream of the inner regions of the galaxy, is made of several
filamentary structures further south. We show five clear filaments in Figure \ref{N4330_filaments} which are highlighted with blue segments and are labelled F1 to F5
from E to W. The position angles (PA) of these filaments, measured anticlockwise from N through E, are 165, 174, 171, 173, 173 deg respectively from F1 to F5.
The direction of the filaments is probably the best indicator of the current direction of the galaxy through the ICM and the average PA = 171 deg indicates that
the galaxy is travelling  almost perfectly towards the north in the Virgo potential. At larger distances downstream of the wind, the filaments appear to shred into 
gaseous blobs which might be a signature of a rapid mixture of the ionised gas with the hot ICM. The direction of motion of \NGAL\ within the hot ICM, as derived
from the PA of the H$\alpha$ filaments, is in good agreement with other determinations from either the numerical simulations of \citet{Vollmer12} or the
estimate from the radio deficit region of \citet{Murphy09}.

\citet{Fumagalli14}, \citet{Fossati16}, and \citet{Consolandi17} presented ram pressure stripped H$\alpha$ tails extending tens of kpc from 
the galaxy disc. Integral field spectroscopic observations revealed that the ionisation of those tails cannot be explained solely by photoionisation.
The line ratios are indeed LINER-like, which indicates turbulence, magneto-hydrodynamic waves, and thermal conduction with the hot ICM as possible
ionisation mechanisms.
In contrast, the tail of \NGAL\ is shorter, more filamentary, and clumpy. These features resemble the magneto-hydrodynamic simulations presented by 
\citet{Ruszkowski14}. These authors find that the stripped filaments can be strongly supported by magnetic pressure and, if this is the case, 
the magnetic fields vectors tend to be aligned with the filaments. Deep spectroscopic observations of the stripped gas revealed by VESTIGE should
be obtained to clarify whether this peculiar morphology is an indication of a different ionisation source, possibly involving strong magnetic fields. 

\section{Stellar population fitting} \label{sec_fitting}
In this section we describe a novel method for accurately reconstructing the quenching history of NGC4330 in various regions along the
major axis by means of stellar population fitting. We make use of two independently developed codes. The first performs a joint 
fitting of spectra and photometry using Monte Carlo techniques to optimally explore the parameter space and the second code fits only
the photometric data points plus the integrated flux in the H$\alpha$ line, as first introduced in \citet{Boselli16a}.
 
 \begin{table}
 \centering
 \begin{tabular}{c c c c}
 \hline
 Region & $R_{\rm maj}~{\rm (kpc)}$ & $L_{\rm maj}~(")$ & $L_{\rm min}~(")$ \\
 \hline
 1  & 10.1 & 55 & 15 \\
 2  & 7.0 & 20 & 15 \\
 3  & 5.3 & 20 & 10 \\
 4  & 3.8 & 15 & 10 \\
 5  & 2.5 & 15 & 5 \\
 6  & 1.2 & 15 & 5 \\
 7  & 0.0 & 15 & 5 \\
 8  & -1.2 & 15 & 5 \\
 9  & -2.5 & 15 & 5 \\
 10 & -3.8 & 15 & 10 \\
 11 & -5.3 & 20 & 10 \\
 12 & -7.0 & 20 & 15 \\
 13 & -10.1 & 55 & 15 \\
 \hline
 \end{tabular}
 \caption{Distance from the galaxy centre measured along the major axis (negative values are towards the east with respect to the photometric centre) and linear sizes for each region in which we extract
 aperture photometry and slit spectra.}
 \label{Table_apertures}
 \end{table}

 \subsection{Aperture photometry} \label{aperturephot}
 
 We divided the emission from NGC4330 into 13 apertures along the major axis, as shown in Figure \ref{N4330_regions} (blue rectangles). The region
 IDs increase from  right to  left. The distance from the centre and the width of each region along the major axis are given in Table \ref{Table_apertures}. 
 In the inner parts of the galaxy there is  a significant structure which is clearly visible in most of the bands. This leads to spatial variation in colour 
 due to the patchy dust extinction. In order to derive a reconstruction of the stellar populations
 which is not biased by these effects we used apertures of 5$"$ width along the minor axis. 
 On the other hand, while moving towards the outskirts of the galaxy we witness a smoother structure perpendicular to the major axis (because the gas and dust
 are almost completely stripped). In these regions we increased the aperture widths to 10$"$ and 15$"$ to preserve a high signal-to-noise ratio 
 with decreasing stellar surface brightness. Since we are mostly interested in deriving stellar population parameters in the outer regions 
 where the quenching of the star formation is complete, the larger apertures ensure that the photometry itself is only marginally affected 
 by beam smearing even in the images with the poorer spatial resolution (FUV, PACS). 
 
 It should also be noted that, due to the high inclination of the galaxy, points perpendicular to the major axis which are at small projected distances 
 from the major axis are physically located at a larger distance. Therefore, by choosing apertures of relatively small width, we ensure 
 that the observed flux  mostly comes from the semi-major axis distance chosen. At large radii this problem becomes less severe; 
 therefore, we can confidently use larger apertures while keeping the contamination from different radii to the minimum.
 
 Aperture photometry is obtained by summing the calibrated flux values in each pixel contributing to a given aperture minus the background value
 in the same aperture. To estimate the background we randomly place each aperture in empty regions of the images and we take the sum of the counts.
 This procedure is repeated 1,000 times;  we then take the median value as the best estimator for the background value. For the UV, optical, and near infrared
 bands, we corrected the measured fluxes for the Galactic attenuation. We assumed the \citet{Fitzpatrick99} Galactic attenuation curve, 
 and a value of $E(B-V)=0.021$ obtained from the dust map by \citet{schlegel98} with the recalibration of \citet{schlafly11} at the position of NGC4330.
 
 The uncertainty on the fluxes
 is obtained as the quadratic sum of the uncertainty on the background (rms of the bootstrap iterations) and the uncertainty on the flux counts. For
 photon-counting devices (all the bands except for PACS), the uncertainty on the flux counts is estimated assuming a Poissonian distribution for the source photons; 
 for PACS images we use the RMS maps distributed by the data reduction pipeline and we estimate the source uncertainty by computing 
 the flux at fixed background level in 100 replications of the flux map, each randomly perturbed with a Gaussian uncertainty taken from the RMS map.
 As a last step, the fluxes from each band are converted into mJy. 
 
 We extract the spectra from the FORS2 reduced 2D spectral images by summing the flux in the spatial pixels contributing to each region along the slit
 direction.
 
 \subsection{Unperturbed and quenched star formation histories}
 In order to reconstruct the quenching history of each radial bin of NGC4330 we need to model its unperturbed radial star formation history (SFH). We use the
 multizone models for the chemical and spectrophotometric unperturbed disc evolution originally presented in \citet{Boissier00} and later updated in 
 \citet{Munoz-Mateos11}. 
 
 
 These models were originally calibrated in the Milky Way in \citet{Boissier99}. They were later generalised to reproduce the properties of nearby spiral galaxies 
 in \citet{Boissier00}). This generalisation was made in the simplest possible way, by introducing scaling relationships inspired by the analytical modes of 
 \citet{Mo98}. As a result, the baryonic mass scales as the rotational velocity $V$ to the power 3, the scale-length as the product of the rotational velocity 
 and the spin parameters ($\lambda$). The star formation law is fixed to that observed in \citet{Boissier03} in the latest version of these models. The 
 accretion history was chosen to reproduce observational trends (early formation of the dense part and of massive galaxies; late accretion for low-density 
 regions and low-mass galaxies).  These models have also been successfully used to interpret the radial light profiles of galaxies in dense environments 
 in order to identify the main quenching mechanism \citep[see e.g.][]{Boselli06a}.
 
 The models have two free parameters: the spin parameter $\lambda$ and rotational velocity $V_{\rm circ}$. The value of the former 
 ranges from $\sim$ 0.02 for compact spiral galaxies galaxies to $\sim$ 0.09 for low surface brightness objects \citep{Boissier00}.
 In this work we adopt $\lambda = 0.05$, consistently with previous works on intermediate mass Virgo galaxies \citep[e.g.][]{Boselli06a}. The rotational velocity
 instead is 120 $\rm km~s^{-1}$ from \HI\ observations \citep{Chung09}. However, this galaxy shows a truncated \HI\ disc; therefore, the measured maximum
 rotational velocity might be lower than the real value. As a test we independently derive the rotational velocity from the baryonic Tully--Fisher relation presented
 by \citet{Lelli16}. Using a stellar mass of $10^{9.8}~{\rm M_\odot}$ and the \HI\ mass-to-stellar mass scaling relation from \citet{Gavazzi13}, we estimate a total
 baryonic mass $M_b = 10^{10.02}~{\rm M_\odot}$, which in turn implies a maximum rotational velocity of 123 $\rm km~s^{-1}$, in good agreement with the
 measured value from \HI\ data.
 
 The SFH model is axisymmetric around the disc rotation axis and is produced in discrete radial bins with roughly 1 kpc step.  
 Therefore, we needed to project it to fit the high inclination of the disc of \NGAL\ and derive the SFH in each of the apertures defined in Section \ref{aperturephot}.
 We assume a width along the minor axis of 5$"$ for each aperture. At large galactocentric radii, where the photometric apertures are larger, we verified that
 there is no clear structure in the flux distribution perpendicular to the major axis. Moreover, at those large radii, the variation in deprojected radius of each pixel contributing
 to the aperture is small (i.e. all the pixels are at large deprojected radii), which motivates our choice of a single aperture for the model extraction.
 
\begin{figure}
\centering
\includegraphics[width=8.5cm]{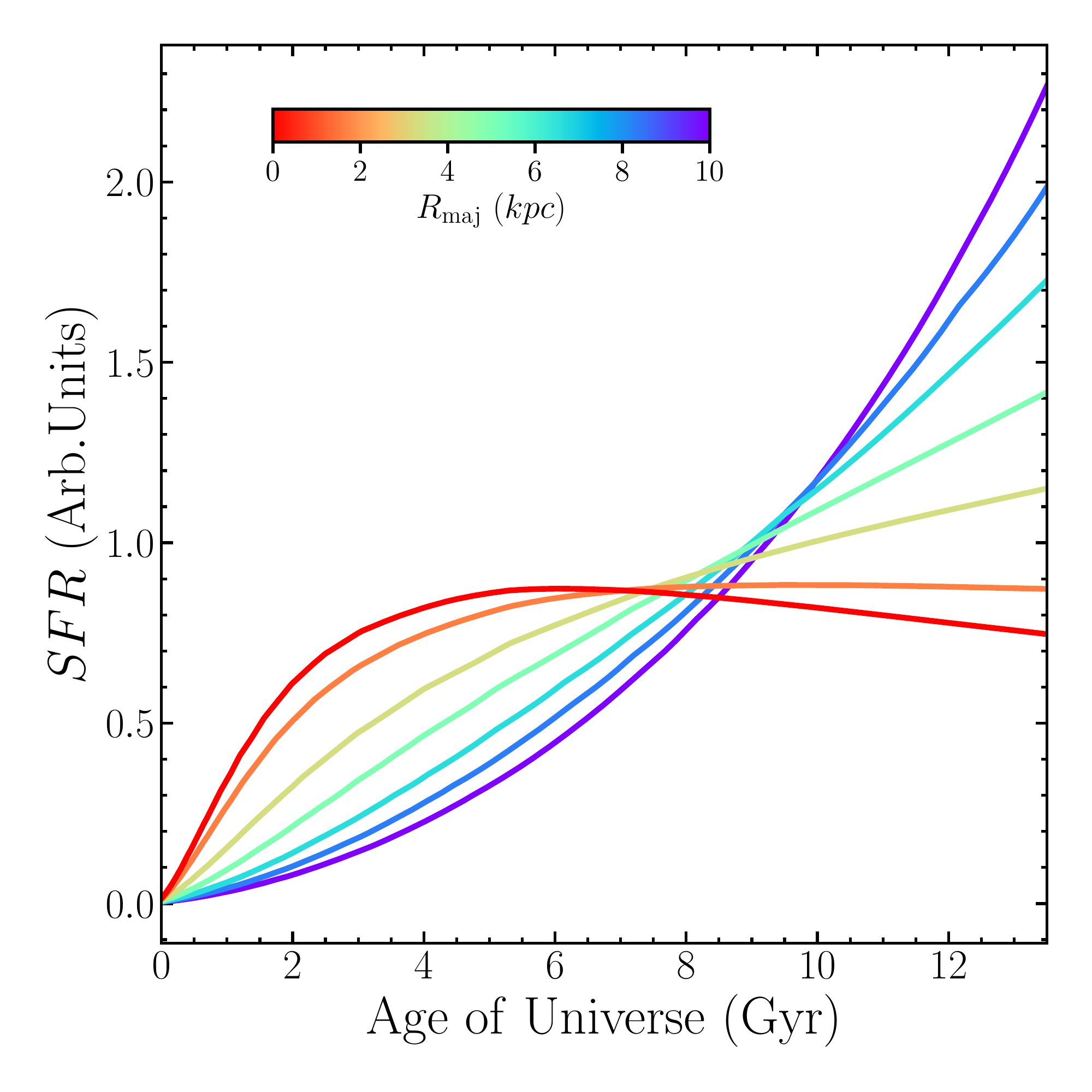}
\caption{Unperturbed star formation histories for regions numbered from 1 (blue) to 7 (red) in Table \ref{Table_apertures}. Each SFH has been normalised to 
produce $1~{\rm M_\odot~yr^{-1}}$ to highlight the variations in shape despite the differences in star formation surface density. 
The colour of each line corresponds to the distance in kpc of each region from the galaxy centre,
as shown by the colour-bar. }
\label{N4330_SFH}
\end{figure}

 The model extraction is performed as follows. First, we interpolate the SFH at each discrete radius to obtain SFHs which run from 0 to 13.5 Gyr in 1 Myr steps.
 Then we define the galaxy as a round disc in pixels of 0.5 x 0.5 kpc, and we populate each bin with a SFH obtained by interpolating the radial bins derived above.
 Finally, we project the model galaxy with the inclination of NGC4330 and we sum the SFHs from each pixel contributing to each of the apertures in order to obtain
 a set of light-weighted SFHs. 
 
 Figure \ref{N4330_SFH} shows the unperturbed star formation histories for regions from 1 (blue) to 7 (red). Each SFH has been 
 normalised to produce $1~{\rm M_\odot~yr^{-1}}$ to highlight the variations in shape despite the differences in star formation surface density. Moving from the 
 galaxy nucleus to the outskirts, the SFH gradually moves to more recent ages. Within this model, the bulk of the mass in the nucleus  formed $5-8$ Gyr 
 ago, while at recent times the growth is mainly taking place at large radii.
 This pattern is consistent with observations of galaxies of similar mass, most notably
 the Milky Way, as inferred from the reconstruction of star formation histories as a function of galactocentric radius from APOGEE spectroscopic maps \citep{Martig16}.
 
 Similarly to \citet{Boselli16a}, we further modify the unperturbed SFHs by introducing an exponential decline in  the SFR to parametrise
 the quenching event. The choice of an exponential quenching pattern is motivated by the fact that while gas is stripped from a given region of the galaxy, the
 star formation activity should decline exponentially for a constant star formation efficiency in order to satisfy the local Kennicutt--Schmidt relation 
 \citep{Schmidt59, Kennicutt98a}. Moreover, this function allows a great range of flexibility and it can model an almost instantaneous and complete quenching event,
 as well as a marginal and gradual suppression of the star formation activity.
 The final star formation histories are defined as
 \begin{equation}
 SFR(t) = \begin{cases} SFR_{\rm unpert.}(t) & \mbox{if } t<t_0-Q_{\rm Age} \\  SFR_{\rm unpert.}(t_0-Q_{\rm Age})\times e^{(t_0-Q_{\rm Age}-t)/\tau_Q}  & \mbox{if } t\ge t_0-Q_{\rm Age}  \end{cases}
 ,\end{equation}
 where $SFR_{\rm unpert.}$ is the star formation history derived above,  $t_0 = 13.5$ Gyr is the current epoch, $Q_{\rm Age}$ is the look-back time for the 
 start of the quenching event, and $\tau_Q$ is the exponential timescale of the quenching. 
 
\begin{figure*}
\centering
\includegraphics[width=17.5cm]{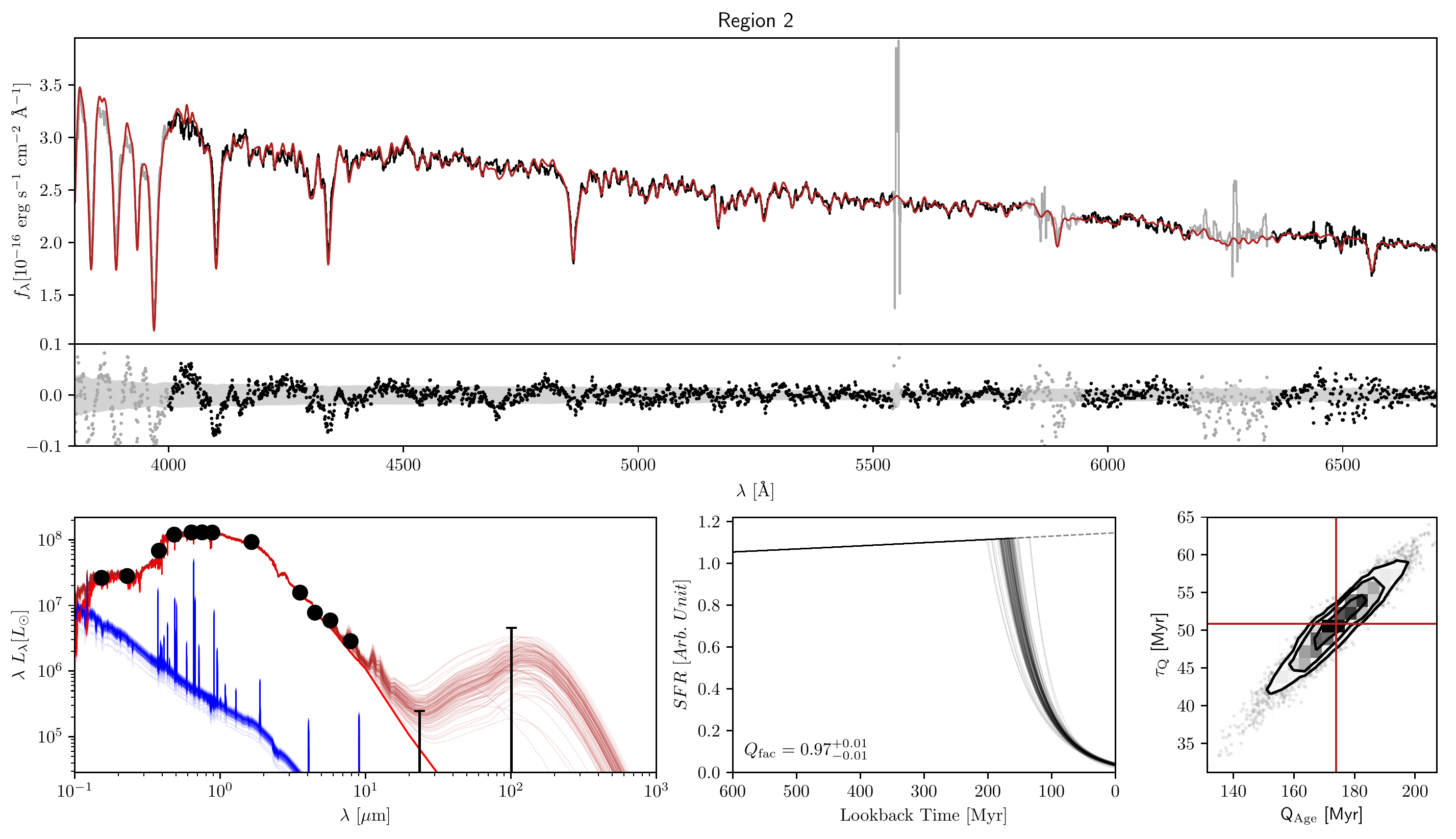}
\caption{Results of the MC-SPF fitting for region 2. Upper panel: FORS2 spectrum (black) and best fit model (dark red). Regions where the spectrum is plotted in grey are not
used in the fit. The fit residuals (Data - Model) are shown below the spectrum and the grey shaded area shows the 1$\sigma$ uncertainties. Lower left panel: Photometric data 
points in black (1$\sigma$ upper limits from non-detections are 
shown as error bars). The blue lines are the stellar emission with nebular lines from the young component (Age<10Myr), the red lines are from the old component (Age>10Myr), and
the dark red lines are the total model including the dust emission. Different lines are obtained by randomly sampling the posterior distribution. Lower middle panel: Reconstructed
SFH from the fitting procedure. Different lines are obtained from the posterior distribution. The grey dashed line shows the evolution of the unperturbed SFH. Lower right panel:
Marginalised likelihood maps for the $Q_{\rm Age}$ and $\tau_Q$ fit parameters. The red lines show the median value for each parameter, while the black contours 
show the 1, 2, and 3 $\sigma$ confidence intervals.}
\label{N4330_MCMC_FIT2}
\end{figure*} 
 
\subsection{Spectro-photometric fitting with MultiNest} \label{MCMCfitter}
One of the methods that we use to reconstruct the quenching histories in NGC4330 is to fit parametric models to the spectro-photometric data using {\sc MultiNest} \citep{Feroz08, Feroz09, Feroz13},
an implementation of the nested sampling algorithm described by \citet{Skilling04}. We  refer to this code hereafter as the Monte Carlo Spectro-Photometric Fitter (MC-SPF). 
We start by constructing a grid of stellar spectra using the quenched SFHs defined above coupled to \citet{Bruzual03} high-resolution models.
Stellar models are linearly interpolated in $Q_{\rm Age}$ and $\tau_Q$ on the fly, and scaled in total luminosity as part of the fitting procedure. 

We added nebular emission lines to the stellar template.
To do so we first computed the number of Lyman continuum ($\lambda<912~\AA$) photons $Q_{H_0}$ for a given stellar spectrum. We then assumed 
that all the ionising radiation is absorbed by the gas, i.e. the escape fraction is zero and that the ionising photons do not contribute to the heating of dust. 
However, we modelled the uncertainty in the conversion of $Q_{H_0}$ into emission line flux by adding a nuisance parameter (${\rm Ly_{scale}}$) in our MC-SPF code. This 
parameter also takes into account differential aperture effects between the 1.3" spectral slit and the larger photometric apertures. 
In summary, the H$\beta$ luminosity can be written as $L_{\rm{H}\beta}=  Q_{H_0} \times {\rm Ly_{scale}} \times 4.55 \times 10^{-13}$ \citep{Osterbrock06}. Finally, we converted the
luminosity into observed H$\beta$ line flux and we obtained the flux in all the other emission lines based on the calculations described by \citet{Byler17}, adopted from 
the Flexible Stellar Population Synthesis code \citep[FSPS;][]{Conroy09}.  We used their photoionisation calculations for $\log(Z/Z_\odot) = 0$ and $\log U = -2.0$, where $Z$ and $U$
are the metallicity and ionisation parameter of the gas respectively, and incorporated the age of the HII regions as an additional free parameter in the fit.  
We  verified that the final derived SFH parameters are relatively insensitive to the choice of metallicity and ionisation parameter.

We assumed a double \citet{Calzetti00} attenuation law to include the effects of dust attenuation on the stellar and nebular emission line spectrum. 
Stars older than 10 Myr and emission lines arising from them are attenuated by a free parameter of the fit ($A_{old}$), while younger stars are attenuated
by $A_{old}$+$A_{extra}$, where $A_{extra}$ is a free parameter and is used to model the extra extinction occurring within the regions of recent star formation.
We coupled the stellar spectrum with dust emission models from the mid- to the far-infrared by means of an energy balance between   
the stellar flux absorbed by the dust and re-emitted in the infrared. We used the models from \citet{Dale02} which add one free parameter in our 
fitting procedure.

We jointly fit the photometric data points and the FORS2 spectra at their native resolution. Multiplicative polynomials are used to remove large-scale shape differences
between the models and the spectra. These differences arise from inaccuracies in the determination of the FORS2 response function, uneven illumination 
across the slit, or imperfect models.  Because the FORS2 response function drops below 4000 $\AA$ we include in the fit only wavelengths above this value. We also
exclude from the fit regions which are significantly affected by bright skylines residuals. The multidimensional likelihood space is sampled 
by using {\sc pyMultiNest} \citep{Buchner14}, a python wrapper for the {\sc MultiNest} code. 

\subsection{CIGALE}
We tested the SFH reconstruction of our new MC-SPF code against the results of the Code Investigating GALaxy Emission \citep[CIGALE;][Boquien et al. in prep.]{Noll09, Boquien16, Ciesla16}.
We made use of the set-up presented in \citet{Boselli16a} and applied to the HRS sample. In brief, CIGALE uses the same 
stellar models described above and coupled with the \citet{Draine14} dust models via the energy balance argument descried above. This code only fits  photometric
data points; however, \citet{Boselli16a} presented a procedure to include the H$\alpha$ observed flux (from spectroscopy or NB imaging) in the fit, 
linking the observed (dust corrected)  H$\alpha$ luminosity to the number of Lyman continuum photons produced in the stellar models. When this additional
constraint is included, the fitting becomes sensitive to the presence (or absence) of the youngest stellar population (age $<10$ Myr), which in turn significantly
improves the SFH reconstruction for recently quenched galaxies \citep{Boselli16a}. These authors further constrained the fit by means of the strength of 
Balmer absorption lines obtained from long-slit spectroscopic observations integrated in narrow bandpasses. In this work, we decided not to use the spectral 
data in the CIGALE fitting, and to constrain the flux from the youngest stars using the dust corrected H$\alpha$ luminosity from VESTIGE narrow-band 
observations, thus to quantify the relative improvement on the quenching age constraints when a full spectral fitting procedure is performed.

\section{Results} \label{sec_results}
We fit the photometry and spectroscopy of all the regions as described in Section \ref{MCMCfitter} in order to accurately quantify the radial quenching history. 
However, as  is clear from Figure \ref{N4330_regions}, for regions from 5 to 9 there is ongoing massive star formation activity. In these regions, the suppression
of star formation due to the effects of ram pressure stripping is hard to quantify and the amount of quenching might only be marginal. 
Moreover, the regions closer to the photometric centre
might not be correctly modelled by the SFH models, due to the possible presence of a pressure supported stellar component (bulge). For these
reasons we limit our analysis to the outer regions. 

\subsection{Results from the MC-SPF fitting code} 

Figure \ref{N4330_MCMC_FIT2} shows the results of the MC-SPF fitting for region 2 using stellar libraries at solar metallicity. In the upper panel the FORS2 
spectrum is shown in black, while the best fit model is overplotted as the dark red line.
The fit residuals (Data - Model) are shown below the spectrum. 
In the lower left panel we show the photometric data points in black (1$\sigma$ upper limits from non-detections are 
shown as error bars) and the total model (stellar + dust emission) is shown in dark red. 
The agreement between model and data is very good both in the high-resolution spectrum and in the photometry from the FUV to the FIR. 
The same quality of fits is achieved for the other regions we include in the analysis. 

\begin{figure*}
\centering
\includegraphics[width=17.5cm]{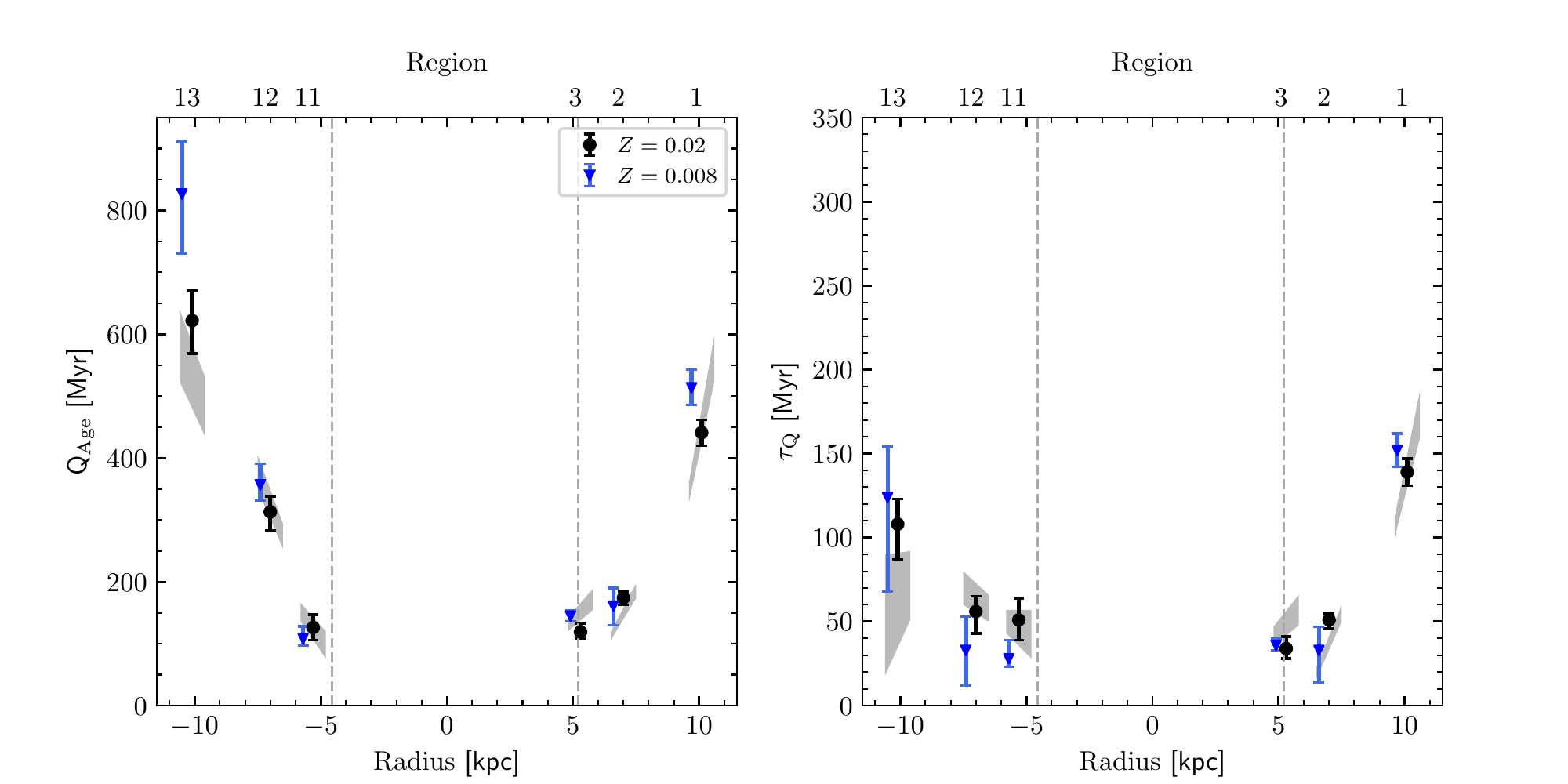}
\caption{Left panel: Derived quenching age ($Q_{\rm Age}$) as a function of galactocentric radius from the marginalisation of the MC-SPF posterior distribution. 
Black points are from stellar libraries at solar metallicity ($Z=0.02$). Blue triangles
are from stellar libraries at subsolar metallicity ($Z=0.008$). The grey error bars show the range of the derived quenching ages for regions shifted by 1/3 of their length
along the major axis. The points belonging to the same region are shifted along the x-axis for clarity. The vertical dashed lines show the truncation radius of the H$\alpha$
emission along the major axis; inside those lines there is ongoing star formation activity. Right panel: same as the left panel, but for the exponential characteristic timescale of the
quenching ($\tau_Q$). }
\label{N4330_MCMC_summary}
\end{figure*}

The lower middle panel shows the reconstructed SFH from the fitting procedure. We define the quenching factor as the amount of 
suppression that is needed to achieve the observed SFR today ($t_0$) compared to that predicted by the unperturbed model:
\begin{equation}
Q_{\rm fac} = 1-(SFR_{pert.}(t_0)/SFR_{unpert.}(t_0))
.\end{equation} 
For region 2, the star formation activity has been recently ($\sim 175$ Myr ago)
truncated on a very short timescale. The quenching is complete $Q_{\rm fac}  \sim 1$ and rapid, although not instantaneous, with the star formation rate being 
suppressed by more than 70\% of its initial value in 50 Myr. Lastly, the lower right panel shows the marginalised likelihood map for the $Q_{\rm Age}$ 
and $\tau_Q$ fit parameters. The red lines show the median value for each parameter, while the black contours show the 1, 2, and 3 $\sigma$ confidence 
intervals. Despite the covariance between $Q_{\rm Age}$ and $\tau_Q$, the SFH after the onset of quenching is very well constrained thanks to the 
high-quality dataset available for \NGAL.

\begin{figure*}
\centering
\includegraphics[width=17.5cm]{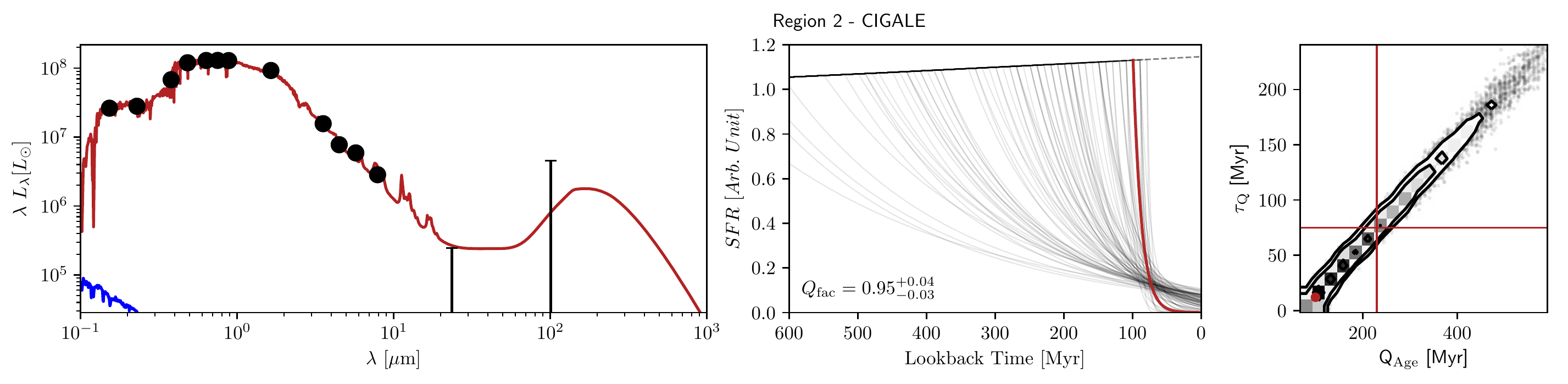}
\caption{Results of the CIGALE fitting for region 2. The panels are the same as the lower panels of Figure \ref{N4330_MCMC_FIT2}, but the fits are 
obtained from the photometric data only plus a constraint on the number of ionising photons coming from the VESTIGE H$\alpha$ imaging 
observations, as first introduced by \citet{Boselli16a}. The red solid lines in the left and middle panels show the CIGALE best fit model and the associated quenching SFH,
respectively. The red lines in the right panel show the median value for each parameter, while the black contours show the 1, 2, and 3 $\sigma$ confidence intervals.
The red dot shows the location of the model with the highest likelihood in the $Q_{\rm Age}$, $\tau_Q$ plane. }
\label{N4330_CIGALE_FIT2}
\end{figure*}

Figure \ref{N4330_MCMC_summary} shows the values of $Q_{\rm Age}$ and $\tau_Q$ from the marginalisation of the MC-SPF posterior distribution 
as a function of galactocentric radius. The black points are from stellar libraries at solar metallicity ($Z=0.02$), while the blue triangles are from stellar 
libraries at subsolar metallicity ($Z=0.008$). The points belonging to the same region are shifted along the x-axis for clarity.  The grey shaded areas 
show the range of the derived quenching ages for regions shifted by 1/3 of their length along the major axis. This allows us to test the robustness 
of the results against the effects of beam smearing of the imaging data and uncertainties on the
radial evolution of the model SFH.   The vertical dashed lines show the 
truncation radius of the H$\alpha$ emission along the major axis; inside these lines there is ongoing star formation activity.

\begin{figure*}
\centering
\includegraphics[width=17.5cm]{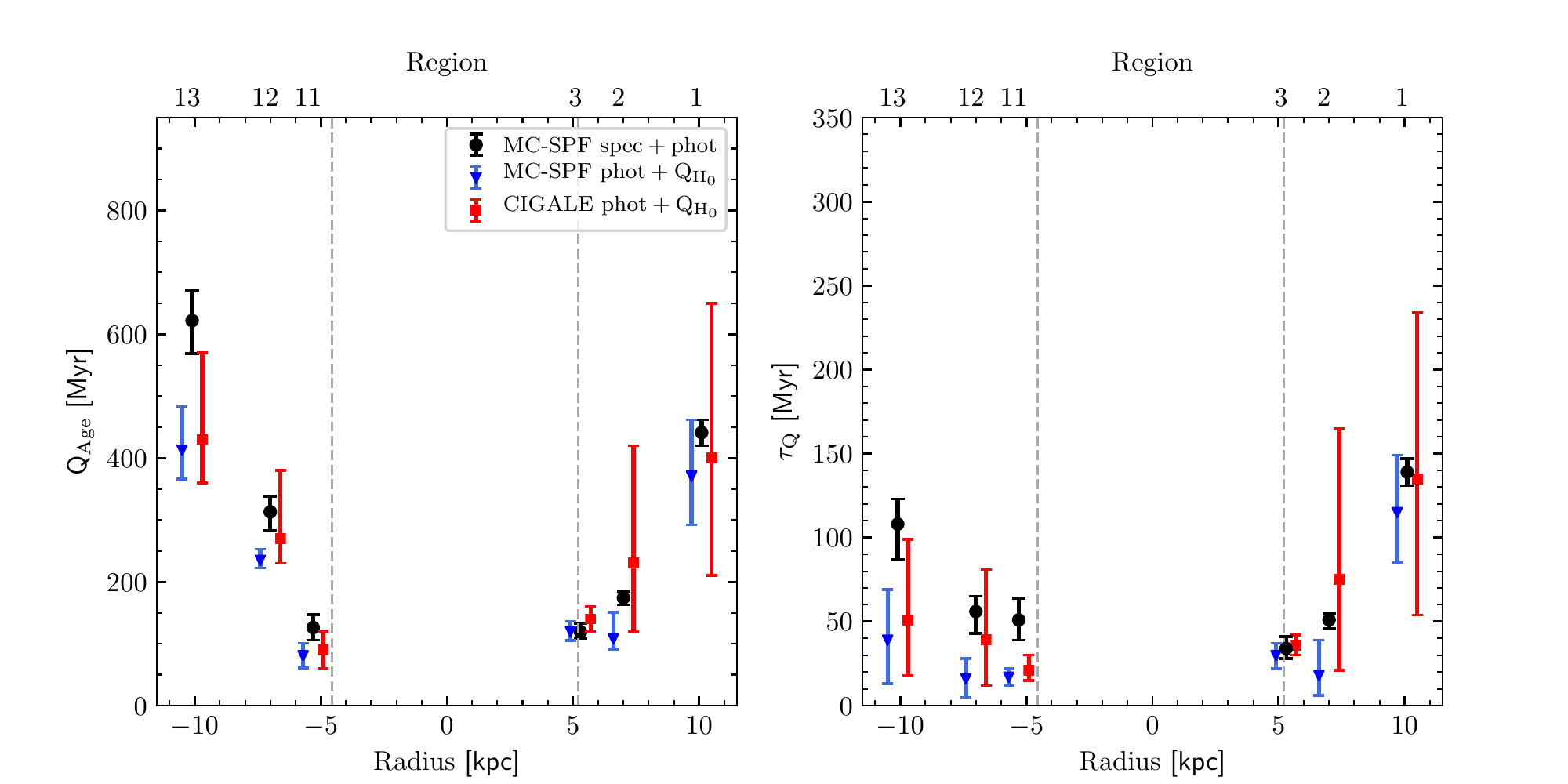}
\caption{Left panel: Comparison of the quenching ages ($Q_{\rm Age}$) as a function of galactocentric radius derived by using different input data into the MC-SPF and 
CIGALE fitting codes. 
Black points are from MC-SPF joint fits to spectra and photometry using stellar libraries at solar metallicity (see also  Figure \ref{N4330_MCMC_summary}). Blue triangles
are from MC-SPF fits to the photometric data plus a constraint on the number of ionising photons coming from the VESTIGE H$\alpha$ imaging 
observations. Red squares are from CIGALE fits and use the same input data as the blue points. The points belonging to the same region are shifted along the x-axis 
for clarity. The vertical dashed lines show the truncation radius of the H$\alpha$ emission along the major axis; inside these lines there is ongoing star formation activity. 
Right panel: Same as the left panel, but for the exponential characteristic timescale of the quenching ($\tau_Q$). }
\label{N4330_CIGALE_summary}
\end{figure*}

The extreme outskirts of the galaxy on the eastern side (the `upturn' region in \citet{Vollmer12}) was quenched $\sim 500-600$ Myr ago. However, 
$Q_{\rm Age}/\tau_Q \sim 4-6$ implying that the quenching event 
was also relatively rapid leaving this region passively evolving for the last $\sim 200-300$ Myr. 
Moving towards the centre of the galaxy, we find an outside-in gradient in the
quenching age which drops to 100 Myr  at the H$\alpha$ truncation radius, again with a relatively fast quenching event. 
On the other side of the galaxy (the `tail' region in \citealt{Vollmer12}), the quenching of the star formation is
more recent ($\sim$ 100--150 Myr ago). 
In these regions the quenching event was also rapid
with 85\% of the star formation activity being suppressed in less than $2\times \tau_Q \sim 100$ Myr.
A notable exception appears at the largest radii in the tail region. There we derive higher values of $Q_{\rm Age}$, 
in combination with longer exponential quenching timescales. The quenching event might then have started at the
same time as on the other side of the galaxy, but then proceeded at a slower rate. In Section \ref{compareCIGALE} we describe
how this result is robustly derived using our fits to the photometry and the spectrum, while a short quenching time cannot
be ruled out using photometric data alone. 

Inside the truncation radius there is a significant amount of ongoing star formation. In these regions the quenching event might not
have started yet or the amount of quenching of the star formation activity might be very small. Therefore, deriving a reliable epoch for the start of the 
quenching is a difficult or ill-posed question. While we obtain reliable fits for region 10 (on the eastern side of the disc), we find that the constraints 
on the quenching history parameters are poor for region 4. Moreover our results are very sensitive to the exact position of the aperture
with a quenching age that ranges from 200 to 800 Myr when moving the region along the major axis by one-third of its size.
This is likely caused by a mismatch between the various photometric bands and the emission detected in the FORS2 slit. 
The high spatial resolution of the CFHT $u$ and $r$ bands (middle panels of Figure \ref{N4330_multiwave})
shows an inhomogeneous dust obscuration pattern as well as clumpy star formation. All these effects are not taken into account into our
fitting procedure, which leads to uncertain fit parameters. Given the caveats affecting these regions, we do not include them in our results.

We also test the effects of stellar metallicity on the quenching history; it has long been known that disc galaxies exhibit on average negative metallicity 
gradients from the centre to the outskirts \citep{Zaritsky94, Rupke10}. This means that we could expect subsolar metallicity in the stripped regions of NGC4330.
Using a stellar library  with Z=0.008 we indeed obtain fits of very similar quality  to those of the solar metallicity models. 
In this case we derive values of  $Q_{\rm Age}$, which are 30\% higher than those obtained with solar metallicity libraries. At fixed age, lower
metallicity stellar populations indeed have a higher optical-UV flux, and an older quenching event is necessary to fit the observed photometry and spectroscopy.
However,  stellar metallicity is not a completely unconstrained parameter. We tested models with Z=0.004 and we obtain poorer fits. 
In these cases, the shape of the models cannot be reconciled with FUV and NUV photometry
by means of an older $Q_{\rm Age}$ and the fit increases the amount of dust extinction (which also reduces the slope of the UV spectrum), revealing tension
with the MIPS and PACS photometric upper limits. We conclude that the metallicity of the outer regions of NGC4330 is between 
the solar value and half of it, and that this degeneracy does not have a significant impact on our results on the quenching of the star formation activity.

\subsection{Results from the CIGALE fitting code} \label{compareCIGALE}

We now compare the results obtained with our new MC-SPT code with those of the CIGALE code, which has been extensively tested and 
characterised in the context of recovering quenching histories by \citet{Boselli16a}.  Because in CIGALE we only use photometric data points (plus a 
constraint on the number of ionising photons coming from the VESTIGE H$\alpha$ imaging observations), we also assess the reduction in the 
uncertainties on the derived parameters when the spectra are included in our MC-SPF code. 

The CIGALE set-up assumes the \citet{Draine14}
dust models, which have more free parameters than the \citet{Dale02} models used in the MC-SPF set-up. With only three photometric points
 sensitive to the shape of the dust spectrum (IRAC 8 $\mu$m, MIPS 24 $\mu$m, and PACS 100 $\mu$m), it is preferable to fit a 
simpler family of models, like those from \citet{Dale02}. However, we are mostly interested in the amount of dust extinction that affects the age sensitive UV and 
optical data, which we constrain from the luminosity of the dust-reprocessed photons. In this respect, the parameters of the dust models are only 
important to the extent that they allow enough freedom to robustly assess the amount of dust extinction. 
We indeed find a very good ($1-2\sigma$) agreement between the values of $A_V$ obtained with the two codes, and we are confident 
that the different dust models are not a source of bias for the optical-UV obscuration and therefore for our stellar population results.

Figure \ref{N4330_CIGALE_FIT2} shows the results of the CIGALE fitting for region 2. The three panels are the same as the lower panels of Figure
\ref{N4330_MCMC_FIT2}. In the left panel, the red line is the best fit CIGALE model, with its associated SFH in the middle panel (red line). The other
lines in the middle panel are obtained by sampling the posterior likelihood space. The right panel shows the marginalised likelihood map for 
the $Q_{\rm Age}$ and $\tau_Q$ fit parameters. The red lines show the median value for each parameter, while the red dot shows the location of the model 
with the highest likelihood. The median values are remarkably close to those obtained
using the MC-SPF code, which indicates that both codes converged to the same quenching history. However, without the spectral information, the 
uncertainty on the derived parameters ($Q_{\rm Age}$ and $\tau_Q$) is much higher, and the degeneracy between them is much stronger. 
The presence of multiple Balmer absorption lines in the spectrum indeed gives us a better sensitivity to intermediate quenching ages (between 100 and
1000 Myr), which is not the case with H$\alpha$ plus UV and optical broad-band data.

Figure \ref{N4330_CIGALE_summary} shows the values of $Q_{\rm Age}$ and $\tau_Q$ and their associated uncertainties as a function of 
galactocentric radius from three different combinations of fitting code and input data. The black points are the same as in 
Figure \ref{N4330_MCMC_summary} and are obtained from the joint photometry and spectrum fits using the MC-SPF code.  The red points are
obtained from CIGALE fits to broad-band photometry plus a  constraint on the number of  ionising photons coming from the 
VESTIGE H$\alpha$ imaging observations as described in \citet{Boselli16a}. We recall that CIGALE computes the 
likelihood in a fixed grid that covers the parameter space. From these values we randomly sample the posterior to extract Monte Carlo-like samples
weighted by the local likelihood. The marginalised distributions in each parameter are then taken from these samples as in the MC-SPF code.  
Lastly, the blue points are obtained from our MC-SPF code, but giving the same input of CIGALE (broad-band photometry and a constraint on 
$Q_{H_0}$ from the H$\alpha$ imaging observations). 
Table \ref{tableres} presents the values of $Q_{\rm Age}$ and $\tau_Q$ and their associated uncertainties for the regions included in the analysis
and with the various fitting methods described in this section.

 \begin{table*}
 \centering
 \begin{tabular}{c c c c c c c}
 \hline
 Region & \multicolumn{2}{c}{MC-SPF (spec+phot)} &  \multicolumn{2}{c}{MC-SPF (phot+$\rm{Q_{H_0}}$)} &  \multicolumn{2}{c}{CIGALE (phot+$\rm{Q_{H_0}}$)} \\
            & $Q_{\rm Age} {\rm [Myr]}$ & $\tau_{\rm Q} {\rm [Myr]}$ & $Q_{\rm Age} {\rm [Myr]}$ & $\tau_{\rm Q} {\rm [Myr]}$ & $Q_{\rm Age} {\rm [Myr]}$ & $\tau_{\rm Q} {\rm [Myr]}$ \\
 \hline
 1   & $441^{+21}_{-21}$ &  $139^{+8}_{-8}$     &  $371^{+91}_{-79}$ &  $115^{+34}_{-30}$  &  $400^{+250}_{-190}$ &  $135^{+99}_{-81}$  \\
 2   & $174^{+11}_{-11}$ &  $51^{+4}_{-5}$        & $108^{+43}_{-17}$ &  $18^{+8}_{-12}$       &  $230^{+90}_{-110}$ &  $75^{+90}_{-54}$  \\
 3   & $119^{+14}_{-11}$ &  $34^{+7}_{-6}$        & $120^{+16}_{-15}$ &  $30^{+7}_{-8}$         &  $140^{+20}_{-20}$ &  $36^{+6}_{-6}$  \\
 11 & $126^{+21}_{-20}$   &  $51^{+13}_{-12}$  & $81^{+20}_{-20}$ &  $17^{+5}_{-5}$           &  $90^{+30}_{-30}$ &  $21^{+9}_{-6}$  \\
 12 & $313^{+25}_{-30}$ &  $56^{+9}_{-13}$      & $235^{+18}_{-13}$ &  $16^{+12}_{-11}$     &  $270^{+90}_{-40}$ &  $39^{+42}_{-27}$  \\
 13 & $622^{+49}_{-53}$ &  $108^{+15}_{-21}$  & $413^{+70}_{-47}$ &  $39^{+30}_{-26}$     &  $430^{+140}_{-70}$ &  $51^{+48}_{-33}$  \\
 \hline
 \end{tabular}
 \caption{Values of $Q_{\rm Age}$ and $\tau_Q$ and their associated uncertainties for regions 1 -- 3 and 11 -- 13 using the joint spectrum+photometry fitting with SPT-MC (Cols. 2 and 3), using the photometry+$\rm{Q_{H_0}}$ fitting with SPT-MC (Cols. 4 and 5), and using the 
 photometry+$\rm{Q_{H_0}}$ fitting with CIGALE (Cols. 6 and 7). All the fits use solar metallicity models ($Z=0.02$).}
 \label{tableres}
 \end{table*}

Despite the excellent photometric coverage 
of the electromagnetic spectrum from the FUV to the FIR, it is immediately clear that both CIGALE and our
MC-SPF code give weaker constraints on the quenching time when using photometry alone, with a significant degeneracy between 
$Q_{\rm Age}$ and $\tau_Q$. This is more noticeable in the tail region where the slope of the UV spectrum is relatively flat 
(see Figure \ref{N4330_MCMC_FIT2}), which combined with the absence of H$\alpha$ emission and dust leaves a large uncertainty 
(100--500 Myr) on the onset of the quenching event. Conversely, 
the inclusion of the spectra significantly improves  the constraints on these parameters because of the inclusion of Balmer absorption lines, 
which are specifically sensitive to these intermediate quenching ages \citep{Poggianti99}. 
We also find a small difference in the uncertainties derived from the MC-SPF and the CIGALE codes: the CIGALE uncertainties are usually 
 larger. This is likely caused by the different fitting approaches. While CIGALE computes the likelihood on a fixed grid, MC-SPF linearly 
interpolates over the input grid to arbitrarily subsample the parameter space around regions of high likelihood, typically resulting 
in smaller uncertainties.

\section{Discussion} \label{sec_discussion}
In the previous section we  show that it is possible to accurately constrain the radial variation of the quenching times in NGC4330,
especially when intermediate-resolution spectroscopic observations are included in the fit. In this section we discuss the implications of
the reconstructed star formation histories in a ram pressure stripping scenario, and more generally for the quenching of galaxies in clusters.

\subsection{Outside-in gradual quenching event}
Figure \ref{N4330_MCMC_summary} clearly shows a radial dependence of the quenching times which is more prominent on the upturn
(eastern) side of the galaxy disc. This behaviour, which is expected by models or numerical simulations of ram pressure stripping \citep{Boselli06, 
Kapferer09, Tonnesen10}, has usually been difficult to characterise, mainly because of the required high sensitivity to the age of the stellar populations. 

\citet{Abramson11} studied the stellar population quenching time in the stripped outer disc of NGC4330 by comparing the observed FUV-NUV
vs. NUV-$r$ colours with predictions from stellar population models. Although the comparison in this parameter space is only qualitative, these authors have
been able to infer a roughly linear increase in the quenching age with radius on the upturn side. Albeit with large uncertainties, their study found an 
increasing gradient of $\sim 300$ Myr moving from from 5 to 8 kpc radius, which is mainly driven by the FUV-NUV colour. This is similar to what we 
find when we restrict our fits to the photometric data alone (see Figure \ref{N4330_CIGALE_summary}, red squares and blue triangles), which 
proves that most of the quenching age sensitivity in photometric data comes from the slope of the UV spectrum. Our best estimates from the
joint fits to photometry and spectra are consistent with the values from \citet{Abramson11}, but with much smaller uncertainties; instead, on the western side the FUV-NUV colour is roughly constant, which these authors interpreted as a constant quenching age out to 8 kpc. Our study confirms
these results, and we do find a recent ($\sim 150$ Myr) quenching age from 5 to 8 kpc. This result, which is visible when using photometric data alone, 
becomes much more robust with the inclusion of the spectrum in the fitting procedure. In this work, with the help of deep spectroscopic data,  
we extend the analysis of the quenching ages to larger radii and we find that at 10 kpc from the centre the quenching age becomes $\sim 400$ Myr.
Taken at face value, it appears that the outer regions of the galaxy, which is where the gas is least gravitationally bound, were quenched $\sim 400-600$
Myr ago. This might have occurred when the galaxy was first suffering from the hydrodynamical interaction of the hot ICM of the Virgo cluster. However,
this assumes that stars on the tail side have lived there for at least this amount of time. While this assumption holds well at smaller radii where the quenching 
times are short, it becomes less appropriate at large radii because of the rotation of the stellar and gas discs. With a maximum rotational velocity of 
120 $\rm km~s^{-1}$ as derived from the \HI\ observations \citep{Chung09}, the rotational period is $\sim 400$ Myr at $R=$10 kpc. This means 
that stars on the upturn side have made at least one full revolution during the stripping event. The old quenching time that we find on the SW side can 
indeed be an effect of the convolution of the quenching events on the upturn side with the galaxy rotation. This would also explain  why the 
exponential timescale is the longest that we have measured in the regions of NGC4330 that we have studied, and that in reality ram pressure stripping has 
only more recently begun to suppress the star formation activity in the SW outskirts region of the galaxy.

With the only exception of the outer regions, we find that the exponential timescales of the quenching are relatively short with an average value of
50 Myr. This means that the SFR is reduced by 85\% in about 100 Myr. These values should be taken as upper limits for the true $\tau_Q$ at a fixed radius
due to the finite (and not small) size of the apertures along the major axis. Even in the limit case of instantaneous quenching at a given radius, 
our apertures  span a range of radii that lead to a non-zero $\tau_Q$. There are other factors that can extend the exponential timescales. First, 
some molecular clouds are too dense to be directly stripped. They become decoupled from the rest of the ISM, and remain for a while in a mostly stripped 
region \citep[e.g.][]{Abramson14}. Star formation in these decoupled clouds limits the evidence of instantaneous quenching. Second, the stellar populations can 
migrate, radially or due to disc rotation, from and into the stripping regions. As described above, this also increases the observed exponential quenching 
timescale above its true value. Despite these caveats, the derived values of $\tau_Q$ are low and at first approximation we can conclude 
that once the gas stripping becomes effective at a given radius, it is followed by a rapid decline in the star formation activity.

\citet{Vollmer12} compared \HI, CO, UV, H$\alpha$, and radio continuum observations to a dynamical model specifically tailored to reproduce 
the morphology of NGC4330 in all these tracers. These authors found that the asymmetry between the upturn and the tail side can be explained by 
the fact that the ram pressure wind is not blowing  onto the galaxy disc face-on. They estimate an inclination angle of $\sim75^\circ$ between the 
galactic plane and the wind direction. This small deviation is enough to produce a radial gradient of the quenching age with radius in the upturn
side of the galaxy. Although the estimated quenching age at the H$\alpha$ truncation radius (95 Myr at $R = 6$ kpc) is consistent with our 
determinations, they find the radial gradient to be relatively small and to reach 160 Myr at $R=10$ kpc. This might be caused by a ram pressure profile
which is too peaked in the dynamical simulation.  When we produce a simulation run with a ram pressure profile which is 3 times broader than the one presented
in \citet{Vollmer12}, indeed we find satisfactory agreement between our reconstructed SFHs and those produced by the simulation in the same
radial bins defined in this work. This can be physically interpreted with a more eccentric orbit that NGC4330 is following across the cluster ICM, compared to 
what was previously assumed. We will present a detailed analysis of these new model runs in a forthcoming paper (Vollmer et al. in preparation).

\subsection{Final fate of NGC4330 in the cluster environment}
The new deep VESTIGE observations coupled with \HI\ maps paint a picture of a multiphase stripping of the galaxy's gas reservoir. 
This is predicted by hydrodynamical simulations \citep{Tonnesen09, Tonnesen10} and naturally arises from the mixing of the galaxy ISM 
with the hot ICM.  Most of the atomic gas missing from spiral galaxies is not recovered in the 
intracluster space \citep{Vollmer07}, which  suggests that the stripped atomic gas must have been transferred to another 
phase. However, the details of the energy transfer, possibly in the presence of magnetic fields, remain poorly understood. In this respect, our detection
of a clear H$\alpha$ tail in \NGAL, combined with the atomic gas tail discovered by \citet{Chung09}, doubles the number of galaxies known
to have tails in these two gas phases in Virgo, with the other object being NGC$~$4388 \citep{oosterloo05}. 
The short length and the clumpy morphology of the H$\alpha$ tail might be an indication of rapid and violent mixing of this gas with the hot ICM. 
However, more detailed models are necessary to understand this observational evidence to their full extent.

Our stellar populations reconstruction shows that the star formation activity in the outer regions of \NGAL\ started to be affected 
by ram pressure around 0.5 Gyr ago. This time is roughly half the cluster crossing time, and given the current position of the galaxy at 0.4 virial
radii, it is plausible to expect that the gas in \NGAL\ started to be removed by ram pressure when the galaxy was first approaching the 
cluster virial radius \citep[see][]{Chung07}. 

The global star formation properties of \NGAL\ and its gas budget also help to understand the current
effects of the interaction with the hot ICM and the final fate of the galaxy. One informative parameter is the \HI\ deficiency: a logarithmic
measurement of the missing amount of \HI\ compared to galaxies of a similar morphological type and 
linear diameter in the field \citep{Haynes84}. For \NGAL\ the \HI\ deficiency ranges from 0.8
\citep{Chung07} to 0.92, according to a more recent determination of the expected \HI\ mass from \citet{Boselli14a}. Those values mean that 84--88\% of the
\HI\ reservoir has been stripped away or converted into another phase. The amount of \HI\ available for star formation might be even lower if the gas 
in the \HI\ tail which is still partially superimposed to the galaxy disc is considered on its way to stripping and is not available for conversion in $\rm H_2$
in the mid-plane of the galactic disc. The $\rm H_2$ deficiency parameter instead ranges from 0.07 to 0.24 if a constant or a luminosity dependent
CO to $\rm H_2$ conversion factor is assumed \citep{Boselli14a}. Both these values are only consistent with  a mild $\rm H_2$ gas stripping. 
Resolved maps of the CO \citep{Lee17} indeed show that the molecular gas distribution closely overlaps  the regions where H$\alpha$ is detected in 
the galaxy disc and star formation activity is currently ongoing ($R<5$ kpc). This evidence suggests a mild decrease in the star formation activity 
of \NGAL\ compared to central galaxies of similar mass. The current star formation rate is $0.18~{\rm M_\odot~yr^{-1}}$  from
a reconstruction of the SFH integrated over the entire galaxy \citep{Boselli16a}. This value is fully consistent with the value of  $SFR =  0.17~{\rm M_\odot~yr^{-1}}$ obtained by 
\citet{Boselli15} by averaging H$\alpha$, FUV, and radio continuum star formation indicators, appropriately corrected for dust absorption and using a Chabrier IMF.
Either of these values places the galaxy 0.6 dex below the main sequence of star forming galaxies in the local Universe, as derived by \citet{Boselli16a},
which is in agreement with an independent determination of the locus of the main sequence galaxies from \citet{Gavazzi15a}.

In summary, \NGAL\ is a partially quenched object which is transitioning from the main sequence to the passive cloud. \citet{Wetzel13} and more
recently \citet{Fossati17} found that satellite galaxies in intermediate gas haloes ($M_h\sim 10^{13}~{\rm M_\odot}$) are quenched on long timescales
of 5--7 Gyr. Because satellite galaxies do not accrete new gas from the cosmic web, this timescale is set by the available gas reservoir at infall into 
a more massive halo. These authors further separated the quenching event into two phases: a delay time ($\sim 2-5$ Gyr), during which the 
star formation rate is unaffected thanks to the replenishment of the molecular gas reservoir from the atomic phase, and after this time, when the star formation 
activity quickly drops due to the reduction of the molecular gas mass via star formation \citep[see e.g.][]{Fumagalli09}. 

In massive haloes 
($M_h> 10^{14}~{\rm M_\odot}$), where ram pressure stripping is effective, the gas consumption timescales are shorter than for field 
galaxies \citep{Boselli14a}. However, these timescales can be longer in cluster galaxies if the star formation rate
slowly decreases over time. The current SFR of \NGAL\ is already significantly below the main sequence of star forming galaxies and an additional drop
in the star formation activity would result in a galaxy that is classified as passive (which however does not imply that the star formation 
activity has ceased completely) by the methods of \citet{Wetzel13}, which are traditionally used to define the environmental quenching timescale.  

From the \HI\ map we estimate that $4.04 \times 10^{8} {\rm M_\odot}$ of atomic gas are currently in the galaxy disc which adds to 
$1.19 \times 10^8 {\rm M_\odot}$ of molecular gas \citep{Lee17}. At the current star formation rate, the two gas phases will be depleted 
in 2 Gyr and 600 Myr, respectively. These numbers are likely to be upper limits. They are only valid if we assume that all the gas can be
converted into stars and that we neglect any additional gas stripping in the future. 
Since \NGAL\ is approaching the dense inner regions of the cluster potential, it will suffer from an even stronger ram pressure 
than the current value \citep{Vollmer12}, thus decreasing the gas budget (and the consumption timescale) as estimated from current 
observations. 

We therefore conclude that both the delay phase and the fading phase are shorter in the presence of ram pressure stripping, and 
that \NGAL\ is likely to have its star formation activity further reduced within the next $\sim2$ Gyr from the 
combined effect of RPS and exhaustion of the residual gas reservoir via star formation. Within this scenario, where the galaxy  drops
by an additional 0.4 dex (a factor $<3$) below the main sequence, it will reach the threshold  of $sSFR<10^{-11}~{\rm yr^{-1}}$
which typically defines passive galaxies in large statistical samples \citep{Franx08, Wetzel13}.

\section{Conclusions} \label{sec_conclusions}
In this work we have performed a detailed reconstruction of the stellar populations in the outer regions of \NGAL, a nearly edge-on 
galaxy currently undergoing ram pressure stripping of its gas reservoir in the Virgo cluster. 

New and deep H$\alpha$ NB observations
taken as part of the VESTIGE large programme revealed a low surface brightness filamentary tail of ionised gas being stripped from the galaxy.
The H$\alpha$ tail, which is $\sim 10$ kpc long,  partially overlaps  a tail of atomic gas, which is also removed from the galactic disc. 
The tail is composed of several parallel 
filaments which clearly indicate the direction of motion of \NGAL\ within the hot ICM. The morphology
of the tail might indicate a significant magnetic field confinement of the stripped gas. However, more detailed spectroscopic investigations are required 
to study the ionisation state of the stripped gas.

We collected new and deep long-slit spectroscopic observations along the disc major axis with VLT/FORS2, and we developed a novel
MC-SPF fitting technique that jointly fits the photometric and spectroscopic observations with stellar population synthesis models.
We dissect the galaxy in radial bins along the major axis and we fit the observations with synthetic star formation histories which are
exponentially truncated to simulate the suppression of star formation due to ram pressure stripping.  The decline is parametrised
by two quantities, the age at which the star formation activity started to decline ($Q_{\rm Age}$) and the exponential characteristic timescale of 
the decline ($\tau_Q$). We find that the outer regions of the galaxy were quenched $\sim 400-600$ Myr ago and that there is
an outside-in gradient of $Q_{\rm Age}$ with radius such that the regions just outside the radius where we detect ongoing star formation 
 started to suffer from quenching $<100$ Myr ago. The exponential timescales we derived are also relatively short (50 Myr, and longer 
only for the outermost regions), which, despite the effect of the finite size of the apertures, is  clear evidence of a rapid suppression of 
star formation once the quenching reaches a given galactocentric radius.

We tested our new code against the results from CIGALE, a photometric spectral energy distribution fitting code which has been  
used to characterise the star formation histories of galaxies in cluster environments \citep{Boselli16a}. We find excellent agreement with
the results of our MC-SPF code when the spectra are not included in the fit, and therefore the two codes are supplied with the same data.
However, the uncertainties on the recovered parameters become significantly smaller when we also fit the spectroscopic data, in
which case we robustly detect a radial gradient of $Q_{\rm Age}$. 

The H$\alpha$ narrow-band imaging observations and our reconstruction of the radial star formation histories clearly indicate the 
ongoing ram pressure stripping of the gas reservoir of \NGAL. This, combined with gas consumption via star formation, will 
eventually lead to the quenching of the star formation activity.

\begin{acknowledgements}
Matteo Fossati thanks Joe Anderson and Henri Boffin for fruitful discussions on the preparation and reduction of the VLT/FORS2 spectroscopic
observations, Francesco Montesano for advice and suggestions on the Python codes, and Laura Paganini for useful suggestions on the 
measurements of the filament position angles. We thank James Taylor for comments
on this manuscript, and the referee Jeffrey Kenney for his comments that greatly improved the quality of the manuscript.
We warmly thank the entire staff of the ESO Paranal Observatory for the excellent support during the observation, and the
Director Discretionary Time Committee for the prompt time allocation.
We are grateful to the whole CFHT QSO team who assisted us in the preparation and in the execution of the observations and 
helped us with the calibration and characterisation of the instrument, allowing us to make the best out of our data: 
Todd Burdullis, Daniel Devost, Billy Mahoney, Nadine Manset, Andreea Petric, Simon Prunet, Kanoa Withington.

Based on observations obtained with MegaPrime/MegaCam, a joint project of CFHT and CEA/IRFU, at the Canada-France-Hawaii 
Telescope (CFHT), which is operated by the National Research Council (NRC) of Canada, the Institut National des Science de 
l'Universe of the Centre National de la Recherche Scientifique (CNRS) of France, and the University of Hawaii.
Based on observations made with ESO Telescopes at the La Silla or Paranal Observatories under programme ID 298.B-5018(A).

AB acknowledges financial support from ``Programme National de Cosmologie and Galaxies'' (PNCG) funded by 
CNRS/INSU-IN2P3-INP, CEA, and CNES, France, and from
``Projet International de Coop\'eration Scientifique'' (PICS) with Canada funded by the CNRS, France.
M. Fumagalli acknowledges support by the Science and Technology Facilities Council (grant number ST/P000541/1).
MS acknowledges support from the NSF grant 1714764 and the Chandra Award GO6-17111X.
MB was supported by the FONDECYT regular project 1170618 and the MINEDUC-UA projects codes ANT 1655 and ANT 1656.
EWP acknowledges support from the National Natural Science Foundation of China through Grant No. 11573002.

This research made use of Astropy, a community-developed core Python package for Astronomy \citep{Astropy-Collaboration13}.

\end{acknowledgements}

\bibliographystyle{aa}
\bibliography{/Users/matteo/Papers/Mypaperlib}

\end{document}